\documentclass[prb, twocolumn,showpacs,preprintnumbers,amsmath,amssymb]{revtex4-1}
\usepackage{graphicx}
\usepackage[citecolor=blue,colorlinks=true,linkcolor=blue]{hyperref}
\usepackage{color}
\usepackage[usenames,dvipsnames]{xcolor}

\begin{document}

\title{Self-learning kinetic Monte Carlo Simulations of Self-diffusion of small Ag clusters on Ag (111) surface}

\author{Syed Islamuddin Shah}
\email{islamuddin@knights.ucf.edu}
\author{Giridhar Nandipati}
\email{giridhar.nandipati@pnnl.gov}
\author{Talat S. Rahman}
\email{talat.rahman@ucf.edu}
\affiliation{Department of Physics, University of Central Florida,  Orlando, FL  32816}

\date{\today}

\begin{abstract}
The self-diffusion of two-dimensional small Ag islands (containing up to $10$ atoms) on Ag(111) surface has been studied using and self-learning kinetic Monte Carlo [J.
Phys.: Condens. Matter 24, 354004 (2012)] simulations. A variety of concerted, multi-atom and single-atom processes were automatically revealed in these simulations. The size dependence of the diffusion coefficients, effective energy barriers as well as key diffusion processes responsible for island diffusion are reported. In addition, we have compared activation barriers for concerted diffusion processes with those obtained from Density Functional Theory (DFT) calculations.
\end{abstract}
\maketitle
\section{Introduction}
Detailed understanding of diffusion processes is essential for various surface related phenomena such as heterogenous catalysis, epitaxial growth, surface reconstruction, phase transition, segregation, sintering and other topics in surface science \cite{PAS, Kax}. Therefore a great deal of effort has been devoted towards the Self-diffusion of single atom and clusters on metal surfaces\cite{diffbook} both theoretically and experimentally \cite{bassett, tsong, wangss, wangprl1, wangprl2, kellog,wen,pai1,giesenss,giesenprl,fern,vander,busse,muller,liu}. There has also been both experimental and theoretical efforts to find the activation barriers and prefactors for adatom diffusion process \cite{Ni1,Ni2,Ni3,Ni4,Ni5,Ni6,Ni7,Ni8,Ni9,Ni10,Ni11,Ni12,Ni13} on various surfaces. Although various theoretical investigations have been performed for Ag/Ag(111) system using embedded atom method (EAM) \cite{Ni4,BoisvertAg111}, corrected effective medium theory (CEM) \cite{SandersAg111}, effective medium theory (EMT) \cite{bruneAg111,Ni6,SandersAg111} and ab initio calculations \cite{BoisvertAg111Abinitio,Ratschstr}, but to the best of our knowledge there has been no systematic experimental or theoretical effort to understand diffusion mechanisms and calculate their activation barriers  for small Ag islands on Ag(111) surface. Accordingly, in this article we focus on the identification and calculation of the diffusion processes and  their activation barriers for self-diffusion of small Ag islands ($1-10$ atoms) on Ag(111) surface as function of island size.

Due to the presence of two three-fold adsorption sites (fcc and hcp) on fcc(111) surface (Fig.~\ref{fig1ab}(b)), adsorption energies for a single atom or cluster on this surface may differ for different systems leading to different diffusion pattern on this surface. It is therefore important to identify these differences in adsorption energies and various activation energies for the diffusion processes in these systems as accuratly as possible. Density functional theory (DFT) based calculations for the activation energies of diffusion processes, are most accurate to resolve such differences but are very time consuming. On the other hand classical molecular static calculations using interaction potential are very fast and easy to manage but their accuracy is determined by the interaction potential used. Therefore in this study we have done detailed DFT calculations for the activation energies for some of the important diffusion processes and compare them with those obtained from interaction-potential based molecular static calculations to validate the accuracy of our approach.

As potential energy surface of the atomically flat fcc(111) surface is least corrugated, it results in low activation barriers even for clusters to diffuse as a whole on this surface. Therefore studies of diffusion processes on this surface is a challenging problem for both experiments and simulations.
For smaller islands (upto tetramer) all possible diffusion processes can be listed, but as island size gets larger it becomes more difficult to enumerate all possible diffusion processes {\it a priori}. On the other hand, as diffusion processes are rare events, molecular dynamics (MD) simulation cannot capture every microscopic processes possible. Therefore, to do a systematic study of small Ag island diffusion on Ag (111) surface we have used on-lattice self-learning kinetic Monte Carlo (SLKMC-II) method \cite{slkmcII}, which automatically finds all the relevant atomistic processes and their activation barriers on-the-fly for system under study.

This paper is organized as follows. In section~\ref{sim_details} we briefly discuss SLKMC-II method and describe the way through which various diffusion processes are identified and their activation barriers are calculated. In section~\ref{results} we present details of single-atom, multi-atom and concerted diffusion processes responsible for the diffusion of Ag islands of size $1-10$ atoms on Ag(111) surface. Section~\ref{Diffusion coefficients and effective energy barriers} provide quantitative analysis of the results of the diffusion coefficients and effective energy barriers of the islands studied. Last section~\ref{Discussion} is devoted to the discussion and conclusions.

\section{Simulation Details}\label{sim_details}
To study self diffusion of Ag islands on fcc Ag(111) surface, we have carried out SLKMC simulations using an extended pattern recognition scheme \cite{slkmcII} which includes both fcc and hcp sites in the identification of neighborhood around an atom. Further details on simulation cell size, calculation of energy barriers for various diffusion processes and other details can be found elsewhere\cite{slkmcII}.

In order to check the accuracy of our approach for the calculation of energy barriers, we have verified the energy barriers of some of the key concerted diffusion processes for various island sizes of Ag/Ag(111) system using climbing image nudged elastic band (CI-NEB) method as implemented in the VASP code \cite{Kresse1,Kresse2} employing the projector augmented wave (PAW) \cite{Kresse2,BlochlPE} and plane-wave basis set methods, setting the kinetic energy cutoff for plane-wave expansion to 500 eV and describing exchange-correlation interaction between electrons by the Perdew-Burke-Ernzerhof functional (PBE) \cite{PBE}. For all cases, we have used slab size of 5x5x4 (for 10 atoms island we use a $7\times 7\times4$ slab) layers with a vaccume of 15 \AA. We relax all surface structures, using the conjugate-gradient algorithm \cite{Numericalrecipes}, until all force components acting on each atom is smaller than 0.01 eV/\AA. In all calculations, after initial relaxation of the slab where all atoms were relaxed, we fix the bottom three layers. In calculating energy barriers and searching transition states, we first use the Nudged Elastic Band (NEB) method \cite{NEB1} for preliminarily determining a minimum energy path, then apply a Climbing Image Nudged Elastic Band (CI-NEB) \cite{CINEB} calculation. We find that using NEB followed by CI-NEB calculations is efficient in searching transition states than performing NEB alone. Here in, we use the term ``(CI-)NEB" to refer a combination of NEB and CI-NEB calculations. In all (CI)-NEB calculations, we used 7 images in each case.

In all our calculations we use the same pre-exponential factor of $10^{12}$ s$^{-1}$ for all diffusion processes for our KMC simulations, this has been proven to be a good assumption for the systems like the present one\cite{handan-prb, handan-ss}.

\section{Results}\label{results}
\subsection{General Details}


In order to uniquely identify whether an adatom is on an fcc or on an hcp site on the fcc(111) surface requires knowledge of at least $2$ layers, namely the top layer of the substrate and the layer below. Therefore, the same convention as in our previous study (Ref.~\onlinecite{slkmcII}) is followed. That is, the center of an upward-pointing triangle formed by the substrate atoms is an fcc site, and the center of  a  downward-pointing triangle is the hcp site . On the fcc(111) surface, all atoms in an island can sit on fcc sites or on hcp sites or on a combination of both sites, called fcc-, hcp- and mixed-island, respectively. Furthermore, activation barriers of single-atom processes depend on various factors including the type of step-edge, namely A- or B- type, (see Fig.1(b) in Ref.~\onlinecite{slkmcII}) along which atom diffuses. Depending on the type of material either the fcc-island or the hcp-island or mixed-island can be energetically favorable, but in case of Ag(111) surface it is always the fcc-island with few exceptions as.

\begin{figure}
\center{\includegraphics [width=5.5cm]{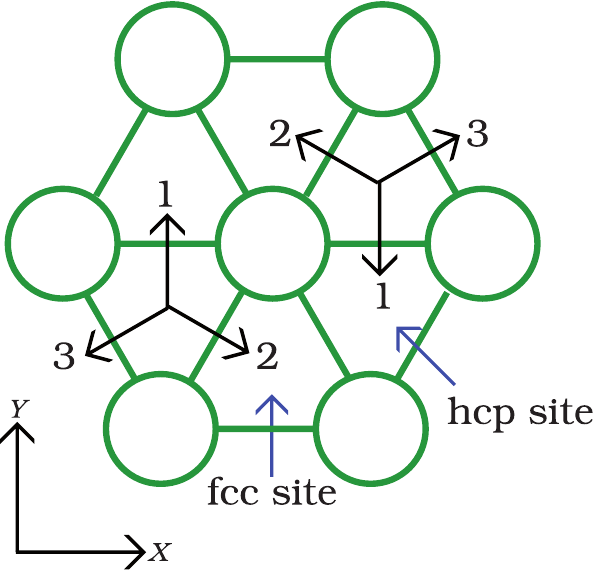} 
\caption{\label{fig1ab}{(fcc and hcp sites on an fcc(111) surface, with corresponding directions for concerted diffusion processes.}}}
\end{figure}

A compact adatom island (called island from hereon) on an fcc(111) surface can move in the three directions (movement of a compact island along other three directions going over the top sites usually results in processes with large energy barriers which are not selected for the range of temperature used in these simulations and hence we do not include them in this study to speed up simulations) shown in Fig.\ref{fig1ab}. Note that the numbering scheme for the directions open to an atom on an fcc site is inverse to that for those open to an atom on an hcp site (see Fig.\ref{fig1ab}). We follow the enumeration convention for directions distinguished in Fig.\ref{fig1ab} throughout the article in tabulating activation barriers for various processes.

Our SLKMC simulations automatically discovered various types of concerted, multi- and single-atom diffusion processes for the island sizes studied. 

Concerted diffusion processes involve all atoms moving together from all-fcc sites to all-hcp sites or vice-versa. A concerted process can be either a concerted translation of the entire island in one of the three directions shown in Fig.~\ref{fig1ab} or concerted rotation of the entire island around the axis through the center of mass of the island perpendicular to the substrate surface, either clockwise or anti-clockwise. Concerted rotational processes may (for example in the case of dimer) or may not (for example in the case of trimer) produce any displacement in the center of mass of the island whereas concerted translation processes always produces the maximum displacement ($a/2\sqrt2$) in the center of mass of the island. Note that in our simulations, concerted rotation processes were observed for dimer and trimer. On the other hand concerted translation processes were observed for compact islands of all sizes. For a particular island size (usually compact islands), activation barriers for the concerted processes along the three directions can be different (depending upon the type of step-edge along the diffusion direction) except for highly symmetric geometries like trimers and heptamers. Also activation barriers for concerted processes for a compact fcc-island along different directions are different than for those of the compact hcp-island of the same size. We will discuss concerted diffusion processes for various island sizes in this section.

Various types of multi-atom diffusion processes like, shearing, reptation and rotation were found mainly for island sizes from 5-10 atoms. In the case of shearing process, part of the island (usually more than one atoms) move from fcc to fcc (or from hcp to hcp sites). On the other hand, reptation~\cite{Chirita} is a two-step shearing process that move the cluster from all-fcc to all-hcp sites or the reverse: first, part of the island moves from fcc to hcp sites (hcp to fcc) creating a dislocation; then the rest of the island moves from fcc to hcp (hcp to fcc). Hence at the intermediate stage, the island has mixed fcc-hcp occupancy. We have also observed  multi-atom rotation processes, mostly involving trimers. Multi-atom processes like shearing is mainly found for compact or almost compact islands where as for mixed-islands, repetition and rotation were observed from island size 5 to 10 atoms. Activation barriers for multi-atom (including shearing, reptation and rotation) processes depend on the type of the island (fcc- or hcp- or mixed-island) as well as type of step-edge along which diffusion process takes place. As island size increases, number of multi- as well as single-atom processes increases. Multi-atom processes are discussed in detail when we take up islands of size 8-10 where as single atom processes are discussed in subsection.~\ref{singleatoms}. 




\subsection{Diffusion Processes}
In this section key diffusion processes of the various general types (concerted, multi-atom and single-atom) are discussed in detail. In addition, activation barriers for some of the concerted diffusion processes are compared with those obtained from DFT calculations. Note that values on top of each pair of figures are the relative energies for those configurations.

\subsubsection{Dimer}
\begin{figure}
\center{\includegraphics [width=8.5cm]{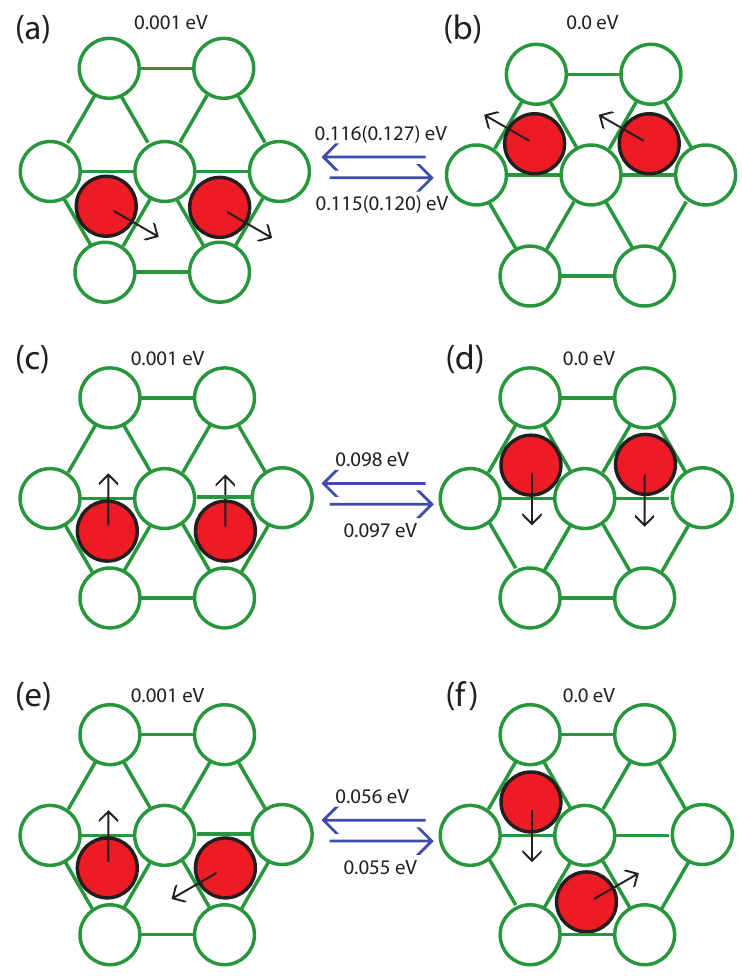} 
\caption{\label{dimer}{Different configurations for dimer along with activation barriers for concerted processes (values in brackets are from DFT calculations). (a \& b) All atoms on hcp sites (hh-dimer) and all atoms on fcc sites (ff-dimer) showing concerted translation processes (c \& d) Same as (a \& b) but showing second type of concerted process. (e \& f) Concerted rotation.}}}
\end{figure}

\begin{figure}
\center{\includegraphics [width=8.5cm]{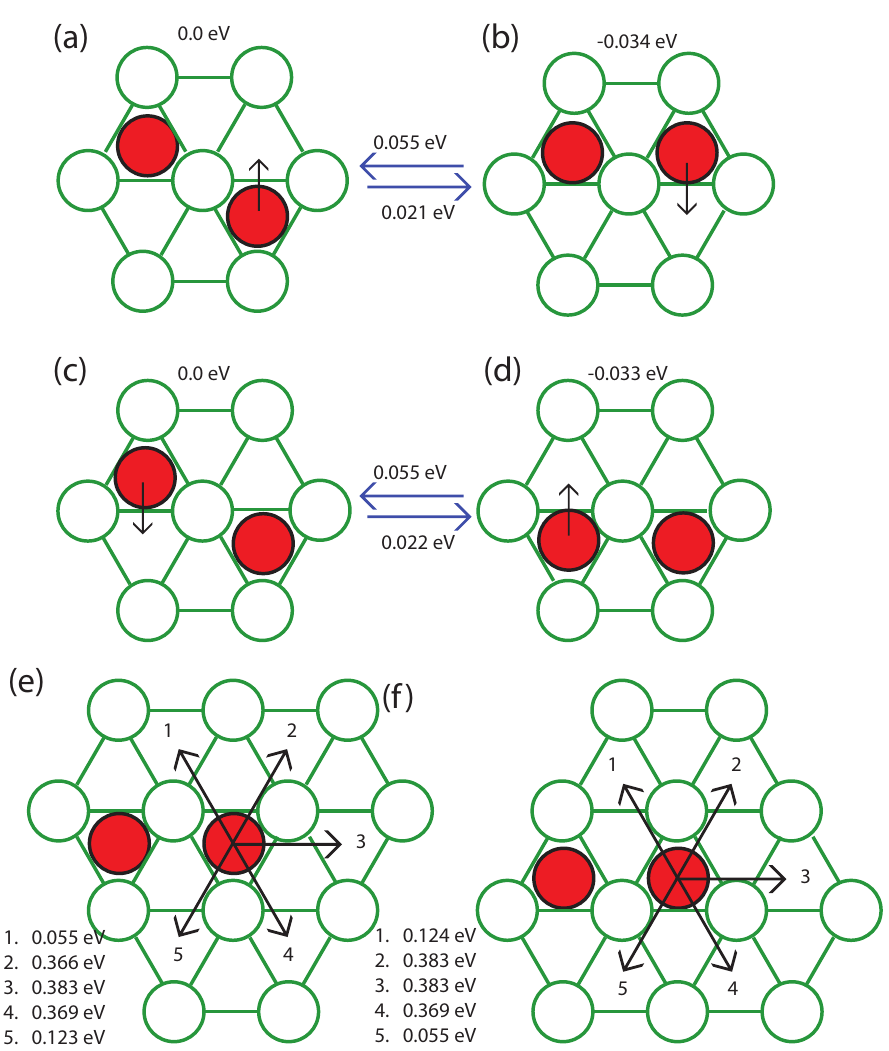} 
\caption{\label{dimer1}{Different configurations for dimer along with activation barriers for single atom processes. (a \& b) Single atom process to transform mixed dimer (a) to ff-dimer and back. (c \& d). Single atom process to transform mixed dimer (c) to hh-dimer and back. (e \& f) Long jump single atom processes including detachment.}}}
\end{figure}

On fcc(111) surface, a dimer can have three possible arrangements; two dimers with both atoms on hcp sites (called hh-dimer, Fig.~\ref{dimer}(a)) or on fcc sites (called ff-dimer, Fig.~\ref{dimer}(b)) and a dimer with one atom on fcc and the other on hcp site (called fh-dimer, Fig.~\ref{dimer1}(a)). Note that both ff- and hh- dimers have equal lenght step edges of both A- and B-type. Our calculations show that ff-dimer is energetically more favorable (0.001 eV from EAM potential and 0.007 eV from DFT calculations) than hh-dimer whereas fh-dimer is less favorable (0.033 eV) than ff-dimer.

There are two different types of concerted diffusion processes, namely concerted translations (two symmetric translations along direction 2 and 3 and another one along direction 1) and concerted rotations (both clockwise and anti-clockwise) for both hh- and ff-dimers as shown in Fig.~\ref{dimer} (a)-(d) and Fig.~\ref{dimer} (e) \& (f), respectively. These processes transforms hh (ff) dimer into ff (hh) dimer. In the case of hh-dimer, activation barrier for concerted translation (Fig.~\ref{dimer}(a)) along direction 2 (Fig.~\ref{fig1ab}(b)) is $0.115$ eV where as along direction 1 (Fig.~\ref{fig1ab}(b)) it is $0.097$ eV (Fig.~\ref{dimer} (c)) while for concerted rotation around the top site ( Fig.~\ref{dimer} (e)) it is $0.055$ eV. For the case of ff-dimer, activation barrier for concerted translation (Fig.~\ref{dimer}(b)) along direction 2 (Fig.~\ref{fig1ab}(b)) is $0.116$ eV where as along direction 1 (Fig.~\ref{fig1ab}(b)) it is $0.098$ eV (Fig.~\ref{dimer} (c)) while for concerted rotation (along direction 2 - anti-clockwise) around the top site ( Fig.~\ref{dimer} (e)) it is $0.056$ eV. Note that, due to symmetry, activation barrier along direction 3 for concerted translation and rotation (clockwise and anti-clockwise) is the same, for both hh- and ff-dimer. 

\begin{table}
\caption{\label{tdimer}Activation barriers (eV) of concerted dimer translations processes. Values in brackets are from DFT. }
\begin{tabular}{ c  c c c c c c }
\hline
\hline
Directions~~&fcc&hcp\\
\hline
\hline
1~~ & 0.098 ~& 0.097 ~\\
2~~ & 0.116(0.127) ~& 0.115(0.120) ~\\
3~~ & 0.116(0.127) ~& 0.115(0.120) ~\\
\hline
\hline
\end{tabular}
\end{table}

Our database also contains single atom processes which converts ff-dimer to fh-dimer (mixed-dimer) with energy barrier of 0.055 eV (Fig.~\ref{dimer1}(b)) with reverse barrier of 0.021 eV (Fig.~\ref{dimer1}(a))to convert back to ff-dimer. Similarly an energy barrier of 0.055 eV converts an hh-dimer (Fig.~\ref{dimer1}(d)) to fh-dimer with a reverse barrier of 0.022 eV (Fig.~\ref{dimer1}(c))to go back to hh-dimer. Our database also contains single atom long jump processes (in which an atom of dimer moves from fcc(hcp) to next fcc(hcp) site) as well as detachment processes as shown in Fig.~\ref{dimer1}(e \& f).
We found that both ff- and hh-dimers diffuse via concerted (both rotaion and translation) as well as single atom processes whereas fh-dimer diffuses via single atom processes. 

\subsubsection{Trimer}

\begin{figure}
\center{\includegraphics [width=4.0cm]{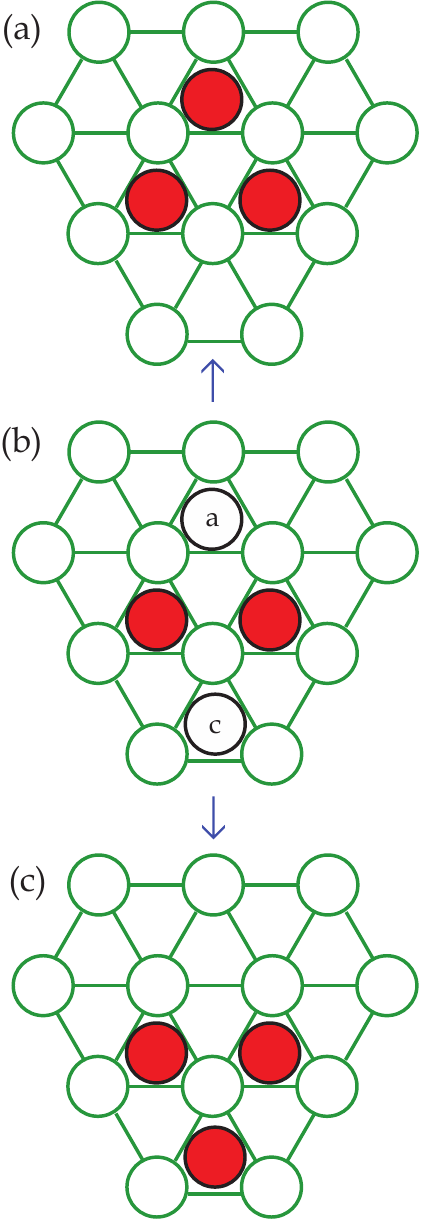} 
\caption{\label{2to3}{Two compact trimers obtained from fcc dimer. (a) F3H, all atoms on fcc sites centered around an hcp site (b) F3T, all atoms on fcc sites centered around a top site.}}}
\end{figure}

\begin{figure}
\center{\includegraphics [width=7.5cm]{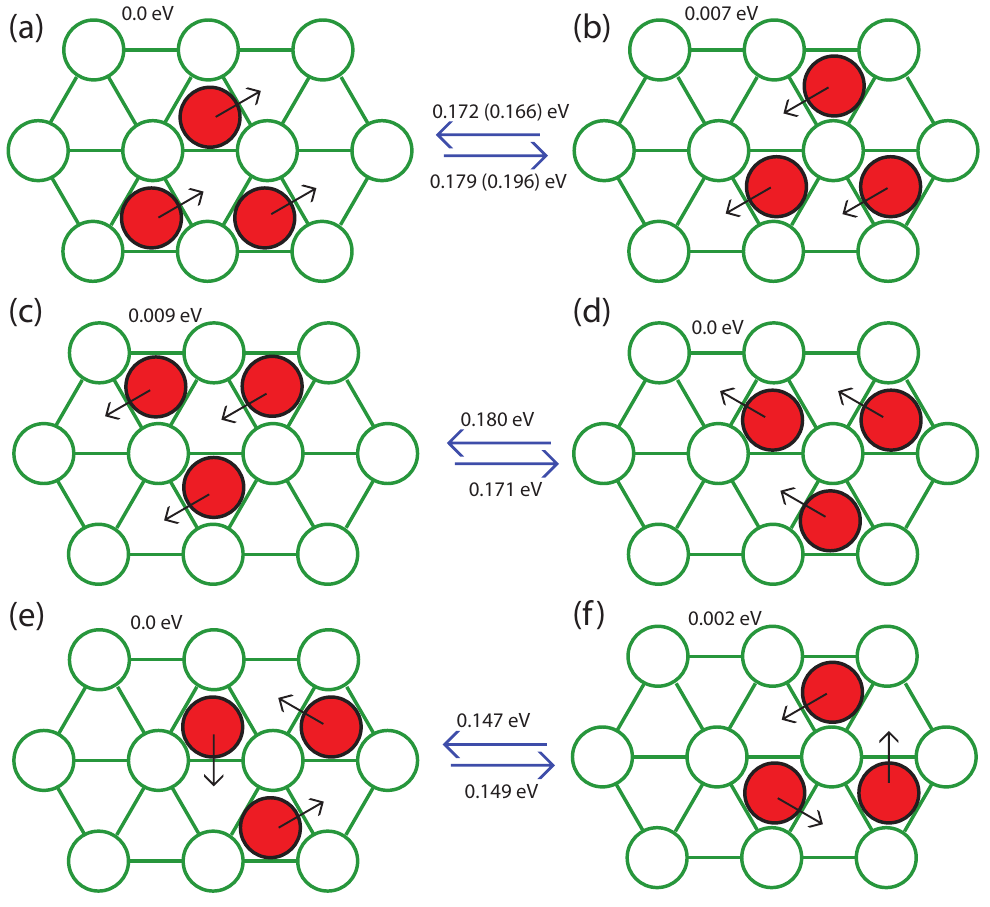} 
\caption{\label{trimer}{Arrangement of atoms in a compact trimer along with concerted processes and their activation barriers. Values in brackets are obtained from DFT based (CI)-NEB calculations. (a) F3H, all atoms on fcc sites centered around an hcp site (b) H3T, all atoms on hcp sites centered around a top site (c) H3F, all atoms on hcp sites centered around an fcc site (d) F3T, all atoms on fcc sites centered around a top site (e) \& (f) concerted rotational process of F3T and H3T and their activation barriers}}}
\end{figure}

Two different types of compact triangular trimers can be obtained from each ff- and hh-dimers of Fig.~\ref{dimer}(b \& a). Two types of compact triangular fcc timers can be obtained from an FF dimer by attaching a third atom to the sites mentioned by black open circles in Fig.~\ref{2to3}(b): one centered around hcp site (called F3H trimer with all step edges of B-type as shown in Fig.~\ref{2to3}(a)) and the other centered around top site (called F3T trimer with all step edges of A-type  as shown in Fig.~\ref{2to3}(c)). Similarly two types of compact triangular hcp trimers can be obtained from HH dimer: one centered around fcc site (called H3F trimer with all step edges of B-type) and the other centered around top site (called H3T trimer with all step edges of A-type) as shown in Fig.~\ref{trimer}(c \& b).

In the case of compact F3T and H3T trimers two types of concerted processes were found, non-diffusive symmetric rotations (both clockwise and anti-clockwise as shown in Fig.~\ref{trimer} (e \& f)) and diffusive translations (symmetric along all three directions) as shown in Fig.~\ref{trimer} (b \& d) along direction 2 for an F3T trimer. Activation barrier for the concerted rotation process for F3T trimer is $0.149$ eV (Fig.~\ref{trimer}(e)) while for H3T trimer it is $0.147$ eV (Fig.~\ref{trimer}(f)). Concerted rotation process transforms an H3T timer into an F3T trimer and vice versa.  In case of a concerted translation process, the activation barrier is $0.180$ eV and $0.172$ eV for F3T and H3T trimer respectively (Fig.~\ref{trimer}(d \& b)). Concerted translation processes transform an F3T(H3T) timer into an H3F(F3H) timer as shown in Fig.~\ref{trimer} (a - d). 

Different than the case with F3T and H3T Trimers, F3H and H3F have only concerted translation processes in all three directions with the same energy barriers (due to symmetry) of $0.179$ eV and $0.171$ eV respectively (Fig.~\ref{trimer}(a \& c)). These concerted diffusion processes transform F3H to a H3T timer and H3F to a F3T trimer. Energy barriers obtained from DFT calculations for the concerted translations for the case of F3H and H3T are 0.196 eV and 0.166 eV respectively which are in close agreement with the values obtained from our static calculations. Our results for the activation energies of various concerted processes agrees very well with those obtained from Effective Medium Theory (EMT) potential as reported in~\cite{Ag_Trimer}.

\begin{figure}
\center{\includegraphics [width=7.5cm]{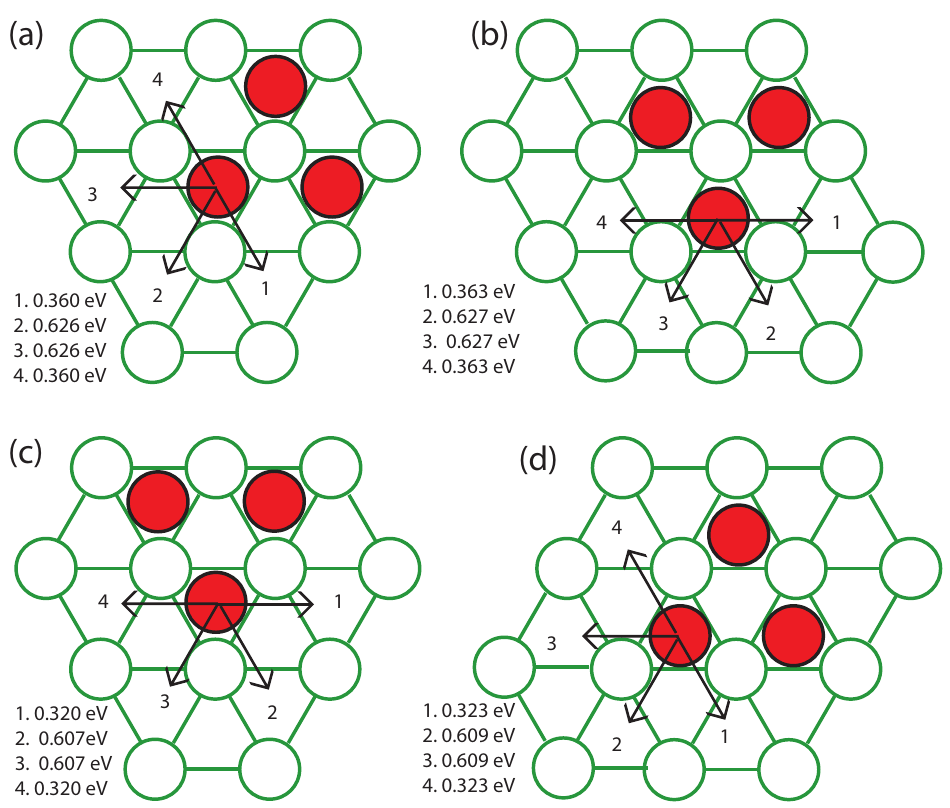} 
\caption{\label{trimer_single}{Single atom processes possible for a trimer and their activation barriers based on type of trimer along different directions.}}}
\end{figure}


Our simulations also finds single atom shape changing processes for trimers. For  compact trimers (Fig.~\ref{trimer_single} (a-d)), an atom can can move in $4$ different directions as shown in Fig.~\ref{trimer_single}. Directions $1$ \& $4$ correspond to edge diffusion process which opens up the trimer (into an non-compact trimer) while direction $2$ \& $3$ correspond to detachment process.  We note that these processes move atoms from fcc (hcp) to nearest fcc (hcp) site.  Activation barriers in these $4$ directions for different types of trimers are shown in Fig.~\ref{trimer_single}. Other types of single atom processes like corner rounding to convert non-compact trimers to linear trimers (and back) are also automatically found in our simulations. We note that as the island size gets bigger, types of these single atom processes becomes large and we categorized them in single atom processes section for cluster sizes lager than $4$ atoms. It can be seen from Fig.~\ref{trimer_single} that single atom processes for trimer have high activation barriers and were rarely observed during KMC simulations.

\subsubsection{Tetramer}

\begin{figure}
\center{\includegraphics [width=8.5cm]{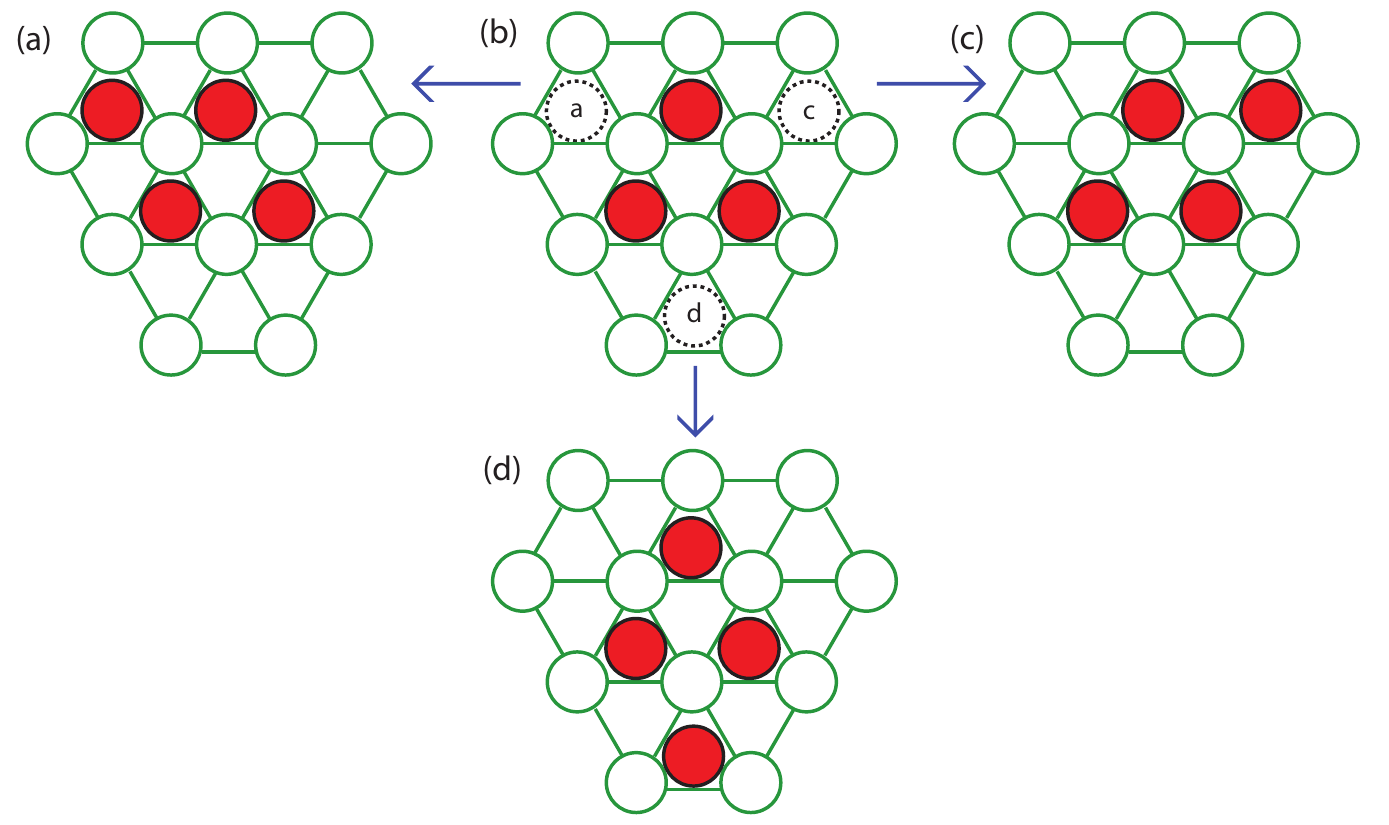} 
\caption{\label{3to4}{Possible shapes of compact tetramers obtained from a compact F3H trimer. (b) Compact F3H tetramer - dotted circles show three possible positions marked as a, b and c for the fourth atom - to get three compact shaped fcc-tetramers. (a \& c) Parallelogram shaped compact fcc-tetramers. (d) Diamond shaped compact fcc-tetramer.}}}
\end{figure}
Adding another atom to three different available positions to the compact F3H trimer shown in Fig.~\ref{3to4} (b) results in the formation of three compact fcc-tetramers with a long (along the line joining the farthest atoms) and a short diagonal (perpendicular to the long diagonal ) as shown in Fig.~\ref{3to4} (a, c \& d). Similar tetramers are obtained from F3T trimer. Note that both F3H and F3T trimers results in the same three compact fcc-tetramers ( Fig.~\ref{3to4} (a, c \& d)) which are symmetric to eachother. Similarly both H3F and H3T trimers results in three compact hcp-tetramers with geometry similar to those of compact fcc-tetramers shown in Fig.~\ref{3to4} (a, c \& d). It should be noted that all compact fcc- and hcp-tetramers have 2 A- and B-steps of same length. 

Our calculations show that fcc-tetramer is 0.002 eV more stable that hcp-tetramer. Three types of concerted processes are possible for a compact tetramer; one along each direction as shown in (Fig.~\ref{fig1ab}(b)). Two out of the three concerted processes (along direction 2 \& 3) for each fcc- and hcp-tetramer are symmetric processes. Our SLKMC simulations also found single atom as well as multi-atom processes for tetramers.
\begin{figure}
\center{\includegraphics [width=8.5cm]{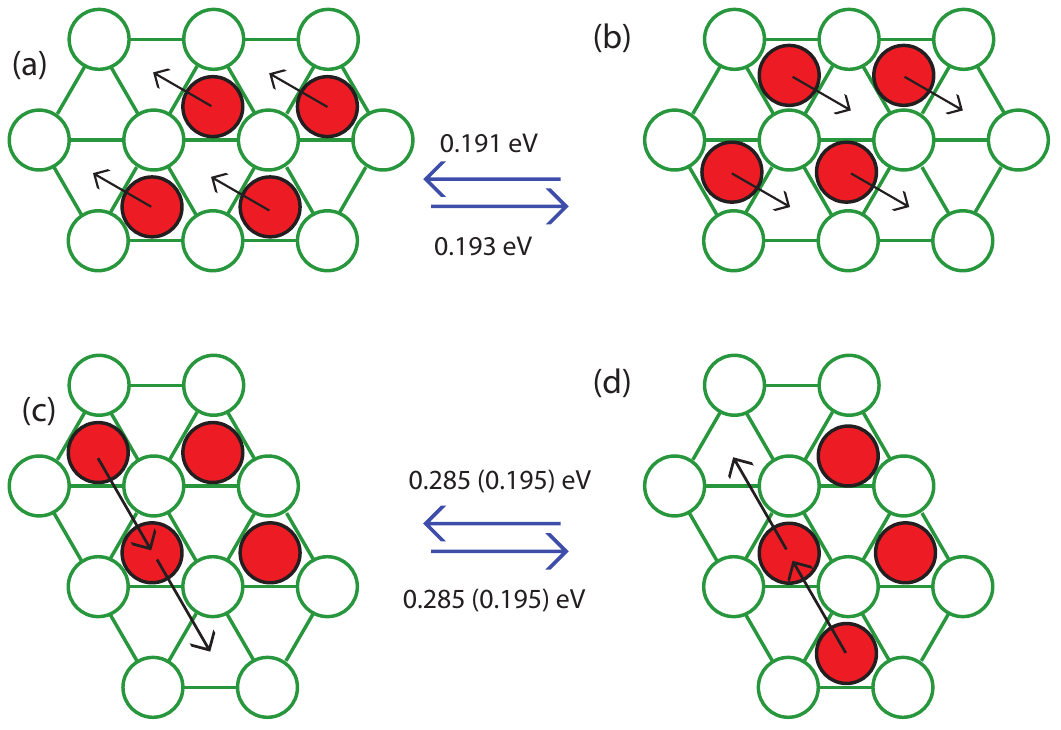} 
\caption{\label{tetramer}{Various diffusion processes for compact tetramers.Values in brackets are obtained from CI-(NEB) based DFT calculations. (a \& b) Concerted diffusion along direction 2 for a compact fcc-tetramer (hcp-tetramer). (c \& d) Multi-atom dimer shearing process for a compact fcc-teramer(hcp-tetramer).}}}
\end{figure}

\begin{table}
\caption{\label{ttetramer}Activation barriers (eV) of concerted processes for compact fcc- and hcp-tetramers as shown in Fig.~\ref{tetramer}(a \& b) }
\begin{tabular}{ c  c c c c c c }
\hline
\hline
Directions~~&fcc&hcp\\
\hline
\hline
1~~ & 0.193 ~& 0.191 ~\\
2~~ & 0.193 ~& 0.191 ~\\
3~~ & 0.245 ~& 0.244 ~\\
\hline
\hline
\end{tabular}
\end{table}


Fig.~\ref{tetramer}(a \& b) shows concerted diffusion process, which converts an fcc-tetramer to an hcp-tetramer, along direction 2 (barrier along direction 3 is also same due to symmetry) for an fcc-tetramer with energy barrier of 0.193 eV with a reverse energy barrier (for an hcp-tetramer to covert to fcc-tetramer) of 0.191 eV whereas energy barrier along direction 1 is 0.245 eV with reverse barrier (for hcp-tetramer) of 0.243 eV (see Table.~\ref{ttetramer}). 
Note that concerted processes in symmetric directions are not shown. We also found multi-atom process as shown in Fig.~\ref{tetramer}(c \& d) in which two atoms (along one of the A-steps in the tetramer) move together in the same direction like a shearing mechanism. This diffusion process has an activation barrier of $0.285$ eV and $0.276$ eV for the compact fcc- and hcp-tetramers respectively. This multi-atom process converts parallelogram shaped fcc-tetramer (Fig.~\ref{tetramer}(c)) to diamond shaped fcc-tetramer (Fig.~\ref{tetramer}(d)) and vice versa. Anothe multi-atom process involving two atoms (along the other A-step of the tetramer in Fig.~\ref{tetramer}(c)), called dimer shearing, converts it into parallelogram shaped fcc-tetramer of Fig.~\ref{tetramer}(a). Similar multi-atom processes are also present for compact hcp-tetramer. These dimer shearing processes are also present in island sizes greater than tetramer. Note that these multi-atom processe are crucial for random motion of tetramer island, their absence results in anomolous diffusion of tetramer. Multi-atom shearing processes along B-steps of the fcc- and hcp-tetramers have very high energy barries and hence we do not report them here.

\subsubsection{Pentamer}

\begin{figure}
\center{\includegraphics [width=8.5cm]{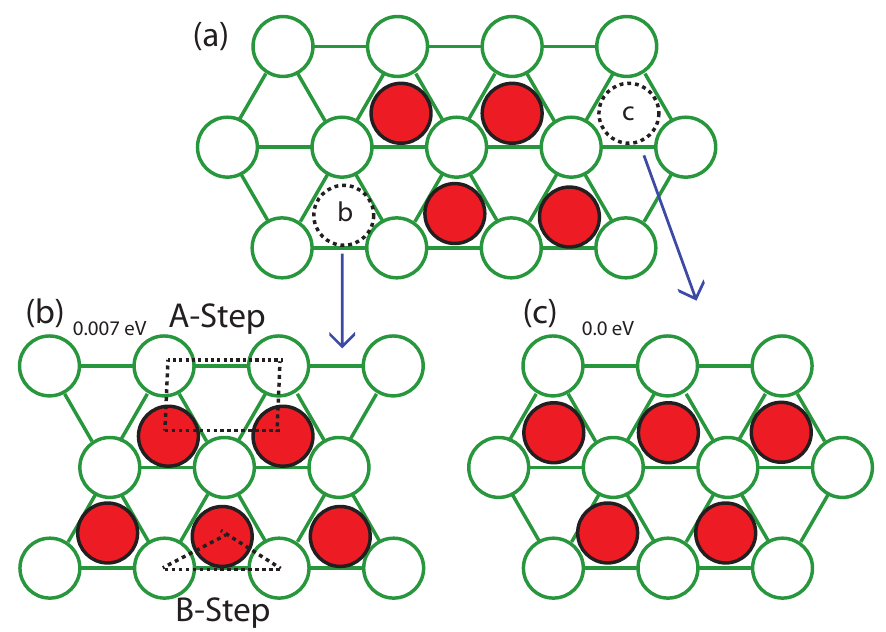} 
\caption{\label{fig5}{Possible shapes of compact fcc-pentamers obtained from a compact fcc-tetramer (a) Compact-fcc tetramer - sites with dotted circles marked as b and c for the fifth atom - to get compact fcc-pentamers: (b) with three B-steps (one B-step edge of length 2 and two B-step edges each of length 1) and one A-step (of length 1). (c) three A-step edges (one A-step edge of length 2 and two A-step edges each of length 1) and one B-step edge (of length 1).}}}
\end{figure}

\begin{figure}[ht]
\center{\includegraphics [width=7.0cm]{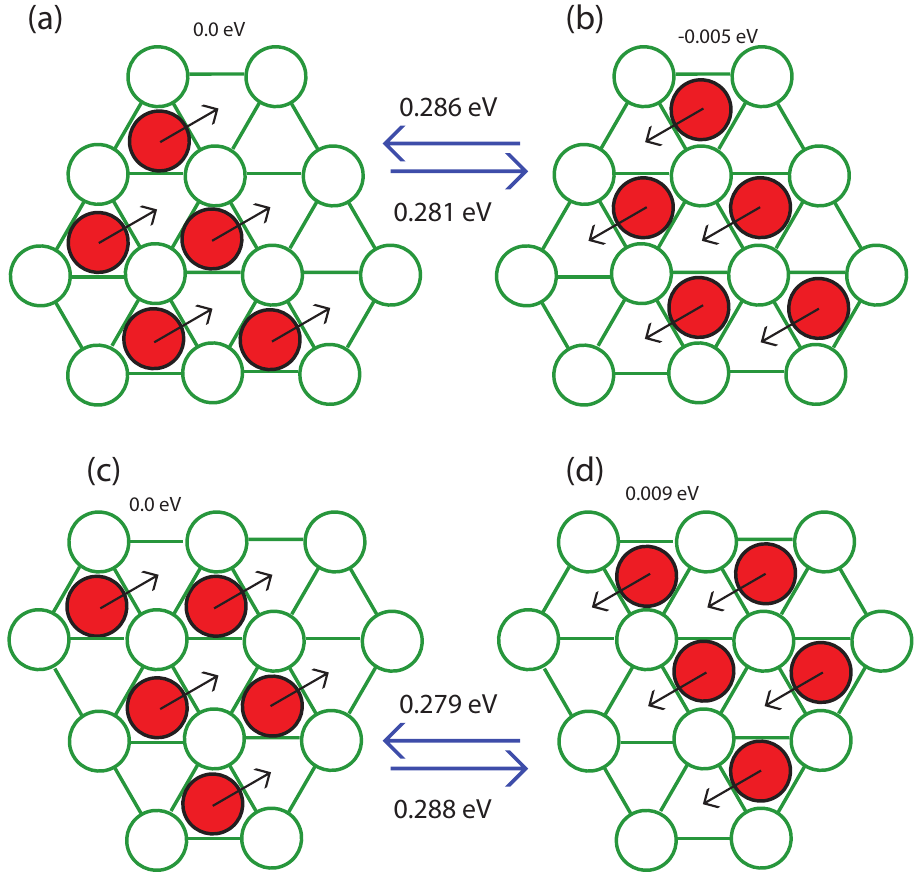} 
\caption{\label{pentamer-m}{Shows concerted diffusion processes for 5-atom compact clusters and their activation barriers}}}
\end{figure}

Two compact fcc-pentamers can be obtained by attaching an atom to the compact fcc-tetramer of Fig.~\ref{tetramer} (a) as shown in Fig.~\ref{fig5}(a). Although the geometry of compact fcc-pentamers thus obtained looks similar, attaching an atom to an A-type step edge of an fcc-tetramer results in an fcc-pentamer with three A-type steps (one A-type step of length 2 and two A-type steps each of length 1) and one B-type step (of length 1) as shown in Fig.~\ref{fig5}(c) (called A-type fcc-pentamer from hereon) whereas attaching an atom to a B-type step edge of an fcc-tetramer results in an fcc-pentamer with three B-type steps (one B-type step of length 2 and two B-type steps each of length 1) and one A-type step (of length 1) as shown in Fig.~\ref{fig5}(b)) (called B-type fcc-pentamer from hereon). Similarly two compact hcp-pentamers called A-type hcp-pentamer (with three A-type steps and one B-type step) and B-type hcp-pentamer(with three B-types and one A-type step) can be obtained by attaching an atom either on A-step or B-step of the hcp-tetramer (see Fig.~\ref{pentamer-m}(b \& d)). Compact A-type fcc-pentamer as shown in Fig.~\ref{pentamer-m}(c)) is 9 meV energetically more favorable than B type hcp-pentamer (Fig.~\ref{pentamer-m}(d))) where as B-type fcc-pentamer (see Fig.~\ref{pentamer-m}(a))) is $5$ meV less favorable and A-type hcp-pentamer (shown in Fig.~\ref{pentamer-m}(b))). 

\begin{figure}[ht]
\center{\includegraphics [width=10.0cm]{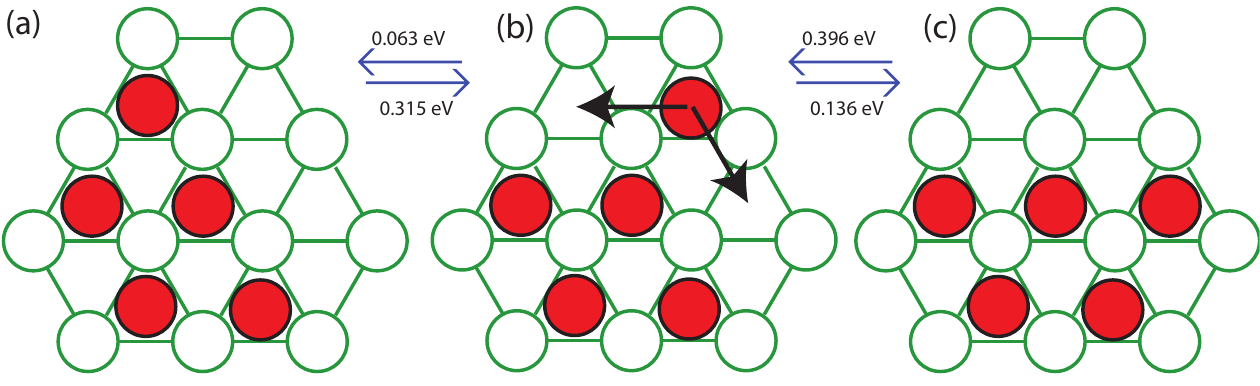} 
\caption{\label{pentamer-single}{Interconvert process for 5-atom compact clusters and their activation barriers. change actual barriers}}}
\end{figure}

In our SLKMC simulation we found that compact A- and B-type fcc- and hcp-pentamers diffuses mostly via concerted diffusion processes which displace the island as a whole from fcc-to-fcc (hcp-to-hcp) and vice-versa. Fig.~\ref{pentamer-m} shows an example of concerted diffusion process and corresponding activation barriers for parallelogram shaped compact A- and B-type hcp- (Fig.~\ref{pentamer-m}(b \& d)) and fcc-pentamers (Fig.~\ref{pentamer-m}(c \& a)) along direction 3. Table~\ref{tpentamer} shows activation barriers for concerted processes for a compact parallelogram shaped A-type and B-type fcc- and hcp-pentamers (see Fig.~\ref{pentamer-m} (a)\&(b)) in all $3$ directions. As can be seen from Fig.~\ref{pentamer-m} (a \& b) that fcc-to-hcp concerted diffusion process for parallelogram shaped B-type fcc-pentamer have lower activation barriers than hcp-to-fcc diffusion process for compact parallelogram shaped  A-type hcp-pentamer as compact parallelogram shaped A-type hcp-pentamer is 5 mEV more stable than compact parallelogram shaped B-type fcc-pentamer. Similarly compact parallelogram shaped A-type fcc-pentamer is 8 meV more stable than compact parallelogram shaped B-type hcp-pentmaer as shown in Fig.~\ref{pentamer-m} (c \& d). Also note that concerted processes in other two directions (not shown in Fig.~\ref{pentamer-m}) are symmetric for both compact parallelogram shaped  A- and B-type fcc- and hcp-pentamers as shown in Table~\ref{tpentamer}. Note that all concerted diffusion processes converts an A-type fcc-pentamer to a B-type hcp-pentamer (Fig.~\ref{pentamer-m} (c \& d)) and vice versa. Similarly, all concerted diffusion processes converts a B-type fcc-pentamer to an A-type hcp-pentamer (Fig.~\ref{pentamer-m} (a \& b)) and vice versa. Hence if we start with an A-type fcc-pentamer, then only B-type hcp-pentamer is accessible through concerted processes whereas B-type fcc- and A-type hcp-pentamers are not accessible. Same is the case when we start either with B-type fcc-pentamer or A-type hcp-pentamer. This results in anamolous diffusion for compact pentamer. In order to get random diffusion for compact pentamer, shape changing single-atom corner rounding processes are cruical. One such example of shape changing single atom corner rounding process is shown in Fig.~\ref{pentamer-single} for an fcc pentamer. To get an A-type fcc-pentamer (Fig.~\ref{pentamer-single}(b \& c)), energy barrier for this single atom corner rounding process is 0.136 eV with a reverse barrier of 0.396 eV while to get a B-type fcc-pentamer (Fig.~\ref{pentamer-single}(b \& a)), it is 0.063 eV with a reverse barrier of 0.315 eV. Similar single atom corner rounding processes to get A- and B-type hcp-pentamers from an hcp-pentamer were also found automatically during simulations. Our database for pentamer also contains various types of multi-atom processes found automatically during simulations.

\begin{table}[ht]
\caption{\label{tpentamer}Activation barriers (in eV) of concerted translation processes for compact A- (and B-type) FCC pentamers and those for compact A- (and B-type) HCP pentamers.}
\begin{tabular}{ c  c c c c c c }
\hline
\hline
Directions~~&fcc & hcp\\
\hline
\hline
1~~ & 0.288 (0.281)~& 0.286(0.279) ~\\
2~~ & 0.290 (0.282)~& 0.287(0.281) ~\\
3~~ & 0.290 (0.282)~& 0.287(0.281) ~\\
\hline
\hline
\end{tabular}
\end{table}

\subsubsection{Hexamer}
\begin{figure}
\center{\includegraphics [width=8.5cm]{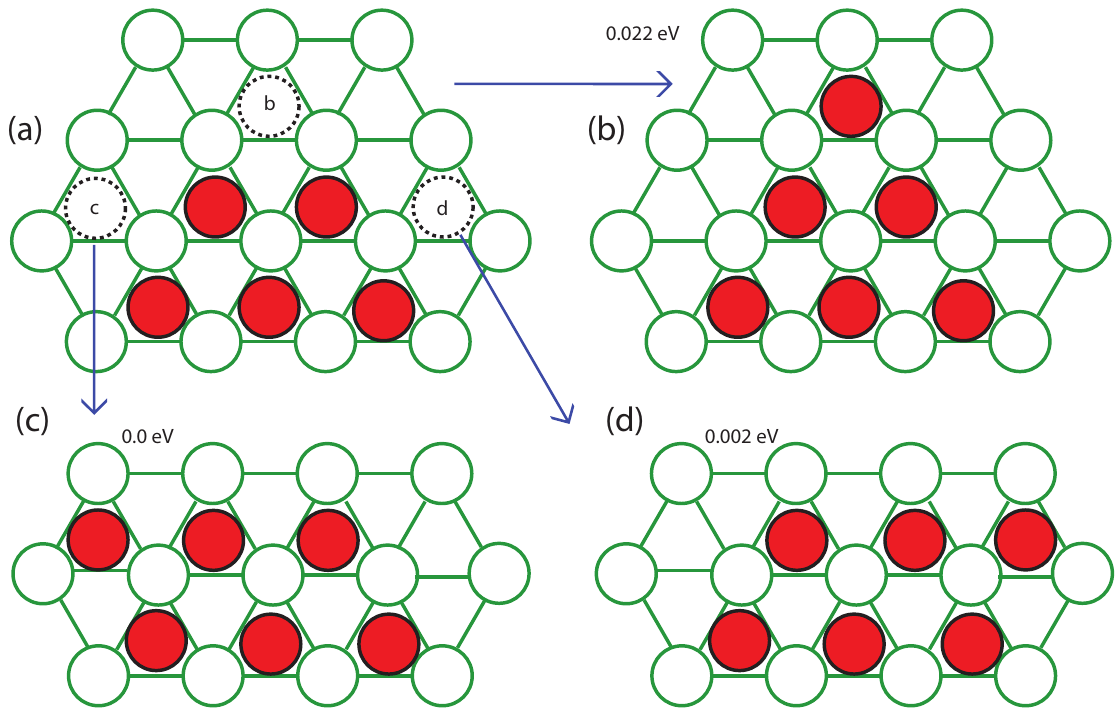} 
\caption{\label{5to6}{Compact hexamer shapes obtained when an atom is added to (a) a B-type fcc-pentamer at three different positions marked by b, c and d. (b) Compact triangular B-type fcc-hexamer with all B-steps. (c) Compact parallelogram shaped fcc-pentamer with same lengths of A- and B-steps, (d) Compact parallelogram shaped fcc-pentamer with same lengths of  A- and B-steps. Note that compact parallelogram shaped hexamers shown in (c) and (d) are symmetric.}}}
\end{figure}

\begin{figure}
\center{\includegraphics [width=8.5cm]{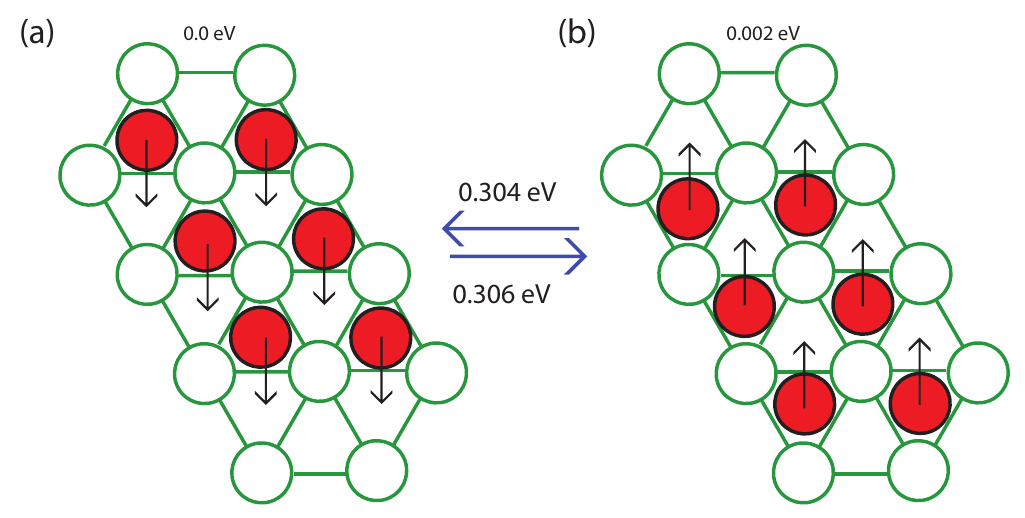} 
\caption{\label{hexamer-1}{Activation barriers along direction 1 for the conceretd diffusion process of a compact parallelogram (a) fcc-hexamer (b) hcp-hexamer.}}}
\end{figure}

\begin{figure}
\center{\includegraphics [width=8.5cm]{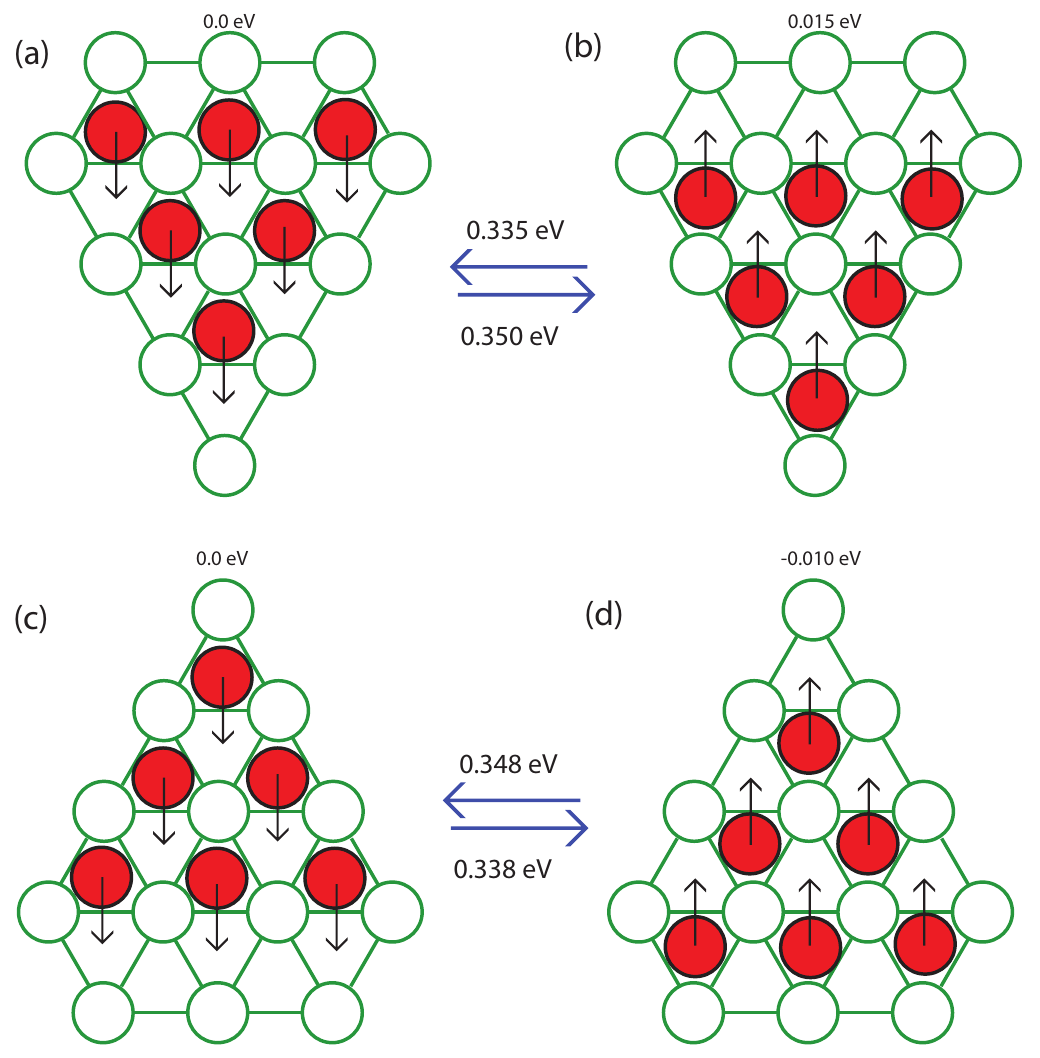} 
\caption{\label{hexamer-3}{Triangular shaped 6-atom island obtained by adding an atom to shorter edge of a 5-atom cluster. (a) all-fcc with B-type step edges (b) all-hcp with A-type step edges (c) all-fcc with A-type step edges (d) all-hcp with B-type step edges. Also shows activation barrier for concerted diffusion in direction $1$.}}}
\end{figure}

Compact B-type hcp-pentamer of Fig.~\ref{fig5}(b) results in three compact hexamers, namely: A triangular B-type fcc-hexamer (with all B-steps each of length 2 as shown in Fig.~\ref{5to6}(b)) and two parallelogram shaped fcc-hexamers (symmetric) with same lengths of A and B-type steps (we will call them parallelogram fcc-hexamer) as shown in Fig.~\ref{5to6}(c \& d). From a compact A-type fcc-pentamer of Fig.~\ref{fig5}(c), three compact hexamers, namely: A triangular A-type fcc-hexamer (with all A-steps each of length 2 as shown in Fig.~\ref{hexamer-3}(a)) and two parallelogram shaped fcc-hexamers (symmetric) with same lengths of A- and B-type steps which are similar to those obtained from compact B-type fcc-pentamer. Similarly, each of the compact A- and B-type hcp-pentamers of Fig.~\ref{pentamer-m}(b \& d) results in two compact triangular hcp-hexamers of A- (with all A-steps as shown in Fig.~\ref{hexamer-3}(d)) and B-type (with all B-steps as shown in Fig.~\ref{hexamer-3}(b)) respectively and two parallelogram shaped compact hcp-hexamers (symmetric) with equal lengths of A- and B-steps (as shown in Fig.~\ref{hexamer-1}(b)).

We found that compact parallelogram fcc-hexamer with similar A- and B-steps (Fig.~\ref{hexamer-1} (a)) is most stable whereas compact parallelogram hcp-hexamer with similar A- and B-steps (Fig.~\ref{hexamer-1} (b)) is 0.002 eV less stable. Also compact triangular A-type fcc-hexamer (Fig.~\ref{hexamer-3} (a)) is 0.015 eV more stable than compact triangular B-type hcp-trangular hexamer (Fig.~\ref{hexamer-3} (b)). 
whereas compact triangular A-type hcp-hexamer (Fig.~\ref{hexamer-3} (d)) is 0.010 eV more stable than compact triangular B-type fcc-hexamer ((Fig.~\ref{hexamer-3} (c))).
Fig.~\ref{hexamer-1} shows concerted diffusion process for compact parallelogram fcc- and hcp-hexamer along direction 1, activation barriers in all three directions are given in Tables.~\ref{thexamer-1}. As can be seen from Table.~\ref{thexamer-1}, the concerted diffusion barrier for compact parallelogram fcc- and hcp-hexamers of Fig.~\ref{hexamer-1} are different in all three directions. Reported in Fig.~\ref{hexamer-3} are the concerted diffusion processes for compact triangular A- and B-type fcc- and hcp-hexamer along direction 1.  Since all three directions for the compact triangular A- and B-type hexamers shown in Fig.~\ref{hexamer-3} are symmetric, so their activation barriers for concerted diffusion are same in all three directions.

\begin{table}
\caption{\label{thexamer-1}Activation barriers (eV) of concerted hexamer translations processes shown in Fig.\ref{hexamer-1} }
\begin{tabular}{ c  c c c c c c }
\hline
\hline
Directions~~&fcc&hcp\\
\hline
\hline
1~~ & 0.306 ~& 0.304 ~\\
2~~ & 0.353 ~& 0.351 ~\\
3~~ & 0.244 ~& 0.242 ~\\
\hline
\hline
\end{tabular}
\end{table}
\subsubsection{Heptamer}
\begin{figure}
\center{\includegraphics [width=7.5cm]{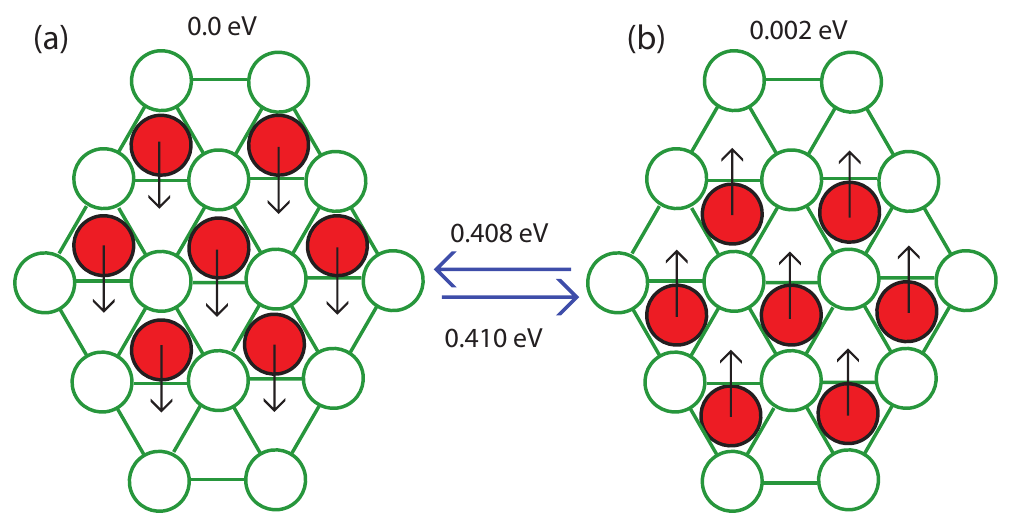} 
\caption{\label{heptamer}{Concerted diffusion processes and their activation barriers in direction $1$ for a (a) fcc-heptamer (b) hcp-heptamer. Note that due to symmetry, activation barriers along direction 2 and 3 are same as those along direction 1 for the two types of heptamer.}}}
\end{figure}

On fcc(111) surface heptamer has a compact closed-shell structure (having same A- and B-type steps of equal lengths) with each edge atom having at least three nearest neighbor bonds as shown in Fig.~\ref{heptamer}. There are two compact hexagonal shaped heptamers, one occupying fcc (called fcc-heptamer) and other occupying hcp (called hcp-heptamer) sites. Our SLKMC simulations found that both fcc- and hcp-heptamer diffuses primarily via concerted diffusion processes, which displace the cluster from fcc-to-hcp and vice versa, and their barriers are shown in Fig.~\ref{heptamer}. This can also be concluded from the effective activation barrier shown in table.~\ref{eff}, which is closer to the average of activation barriers shown in Fig.~\ref{heptamer}. Since the compact heptamer has a symmetric shape accordingly activation barriers in all the other directions are also the same as shown in Fig.~\ref{heptamer}. We note that fcc heptamer is $0.002$ eV more favorable than hcp heptamer. Our database for hepatmer also contains various single- and multi-atom processes as well.

\subsubsection{Octamer}

\begin{figure}
\center{\includegraphics [width=7.0cm]{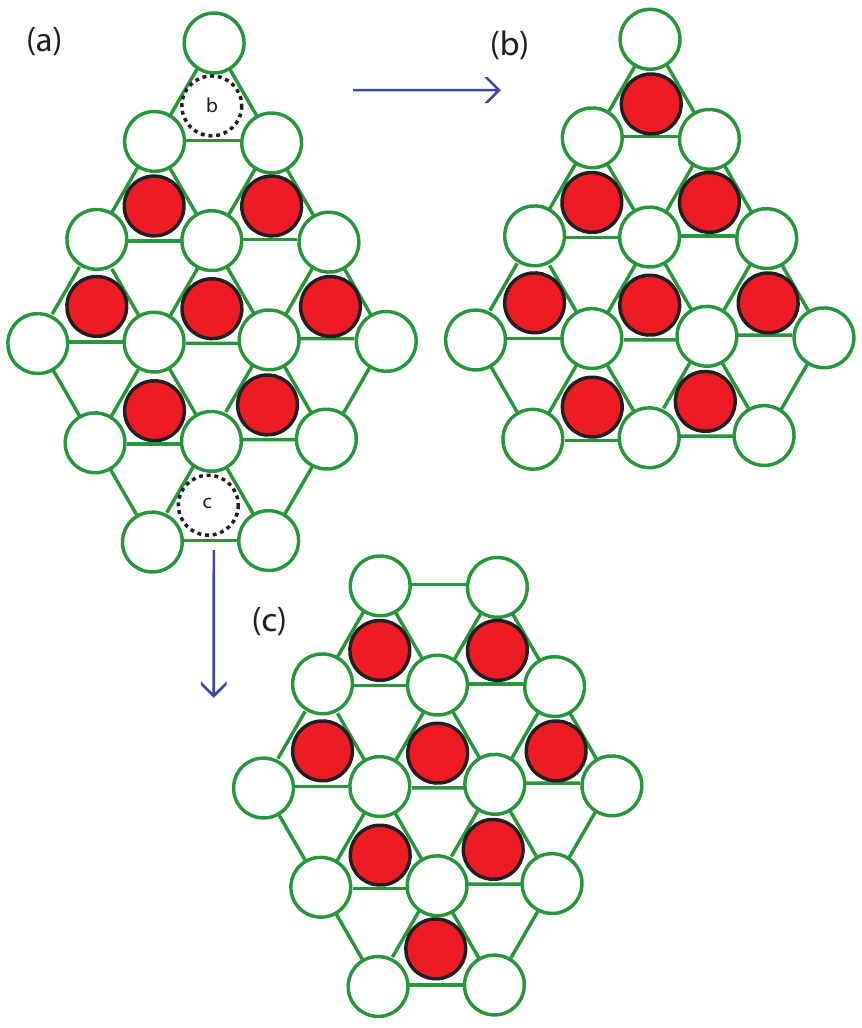} 
\caption{\label{7to8atoms}{Compact geometries for fcc-octamers obtained by attaching an atom to (a) a compact fcc-heptamer at positions marked as b and c with dotted circles. (b) fcc-octamer with overall long B-type step edges than A-type step edges (called B-type fcc-octamer). (c)fcc-octamer with overall long A-type step edges than B-type step edges (called A-type fcc-octamer).}}}
\end{figure}

\begin{figure}
\center{\includegraphics [width=7.0cm]{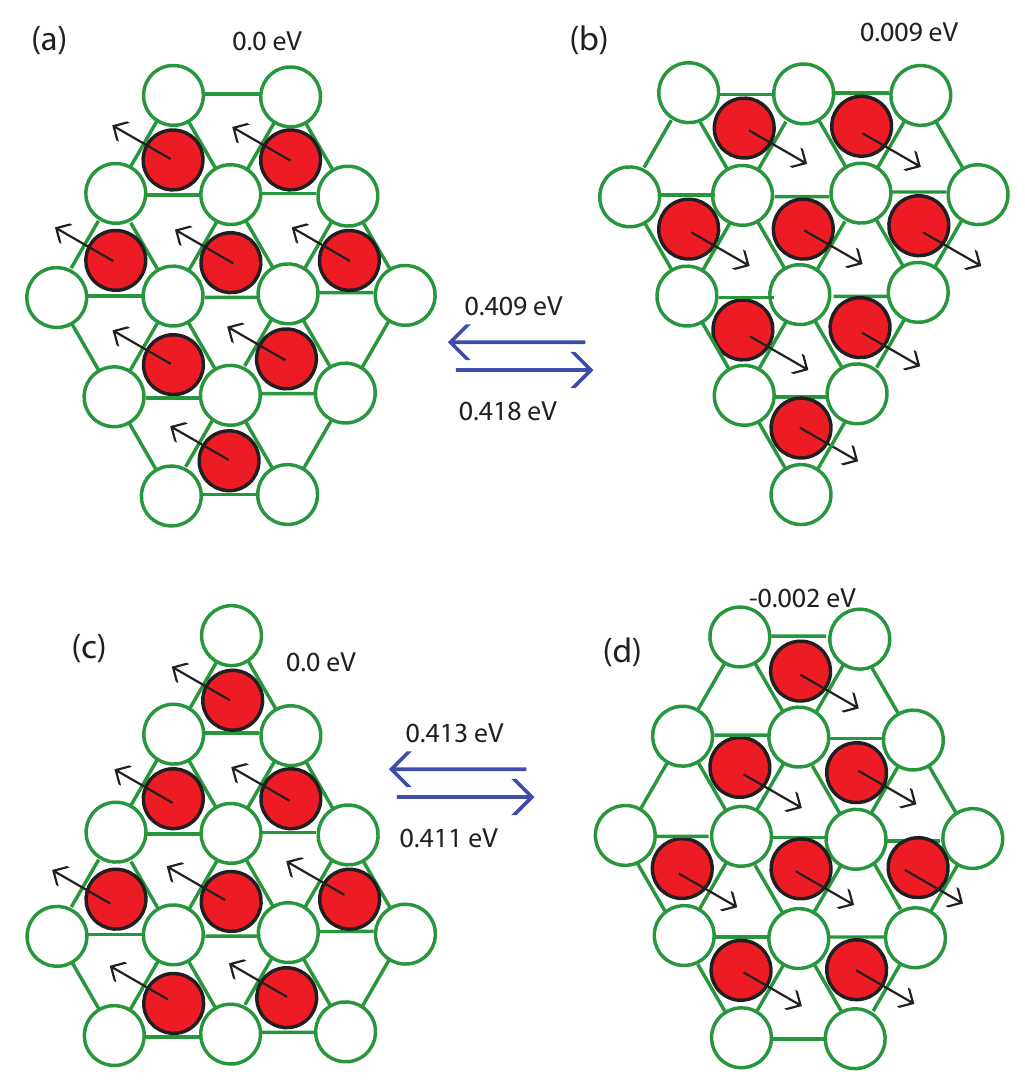} 
\caption{\label{8atom-1}{Concerted diffusion processes along direction 2 for a compact (a) A-type FCC octamer (b) B-type HCP octamer (d) B-type FCC octamer and (d) A-type HCP octamer.}}}
\end{figure}

Adding another atom to the compact fcc-heptamer, we get two compact octamers, one with overall long B-type step egdes than A-type step edges (called B-type fcc-octamer) obtained by attaching an atom to one of the A-steps of fcc-heptamer (Fig.~\ref{7to8atoms}(a \& b)) and another one with overall long A-type step egdes than B-type step edges (called A-type fcc-octamer) obtained by attaching an atom to one of the B-type step edges of fcc-heptamer (Fig.~\ref{7to8atoms}(a \& c)). Similarly two compact B- and A-type hcp-octamer can be obtained by attaching another atom to any of the A- and B-type step edges of the compact hcp-heptamer of Fig.~\ref{heptamer} are shown in Fig.~\ref{8atom-1} (b \& d). Compact A-type fcc-octamer is 0.008 eV more stable than compact B-type hcp-octamer whereas compact A-type hcp-octamer is 0.002 eV more stable than compact B-type hcp-octamer.

Compact octamers diffuse via concerted diffusion processes which displaces the island from all-fcc sites to all-hcp sites and vice versa as shown in Fig~\ref{8atom-1} (a - d). In general, activation barriers for compact A- and B-type fcc-octamers are different. Same is the case for A- and B-type hcp-octamers. It can be seen in the Fig.~\ref{8atom-1} (a - d) that concerted diffusion process convert an A-type fcc-octamer into a B-type hcp-octamerand vice versa.
 Table.~\ref{8atom} gives activation barriers for concerted diffusion processes in all $3$ directions for both compact fcc- and hcp-octamers (see Fig.~\ref{8atom-1}). We note that concerted diffusion processes in the directions $2$ and $3$ are symmetric and have lowest energy barrier whereas for direction 1 it is highest.

Although compact octamers diffuse primarily via concerted diffusion processes, we found in our simulations that both multi-atom (including shearing, reptation, rotation etc) and single-atom (including kink attachment/detachment, corner rounding etc.) processes are also relatively common. Accordingly we will discuss multi-atom processes particular to this island size here, while single atom processes, which are common to all island sizes, will be discussed in detail later.

\begin{table}
\caption{\label{8atom}Activation barriers (eV) for concerted diffusion processes for compact A-type (B-type) fcc-octamers and compact B-type (A-type) hcp-octamers as shown in Fig.~\ref{8atom-1} (a - d).}
\begin{tabular}{ c  c c c c c c }
\hline
\hline
Directions~~&fcc & hcp\\
\hline
\hline
1~~ & 0.472 (0.467) ~& 0.463 (0.469)~\\
2~~ & 0.418 (0.411)~& 0.409 (0.413)~\\
3~~ & 0.418 (0.411)~& 0.409 (0.413)~\\
\hline
\hline
\end{tabular}
\end{table}

\begin{figure}
\center{\includegraphics [width=8.5cm]{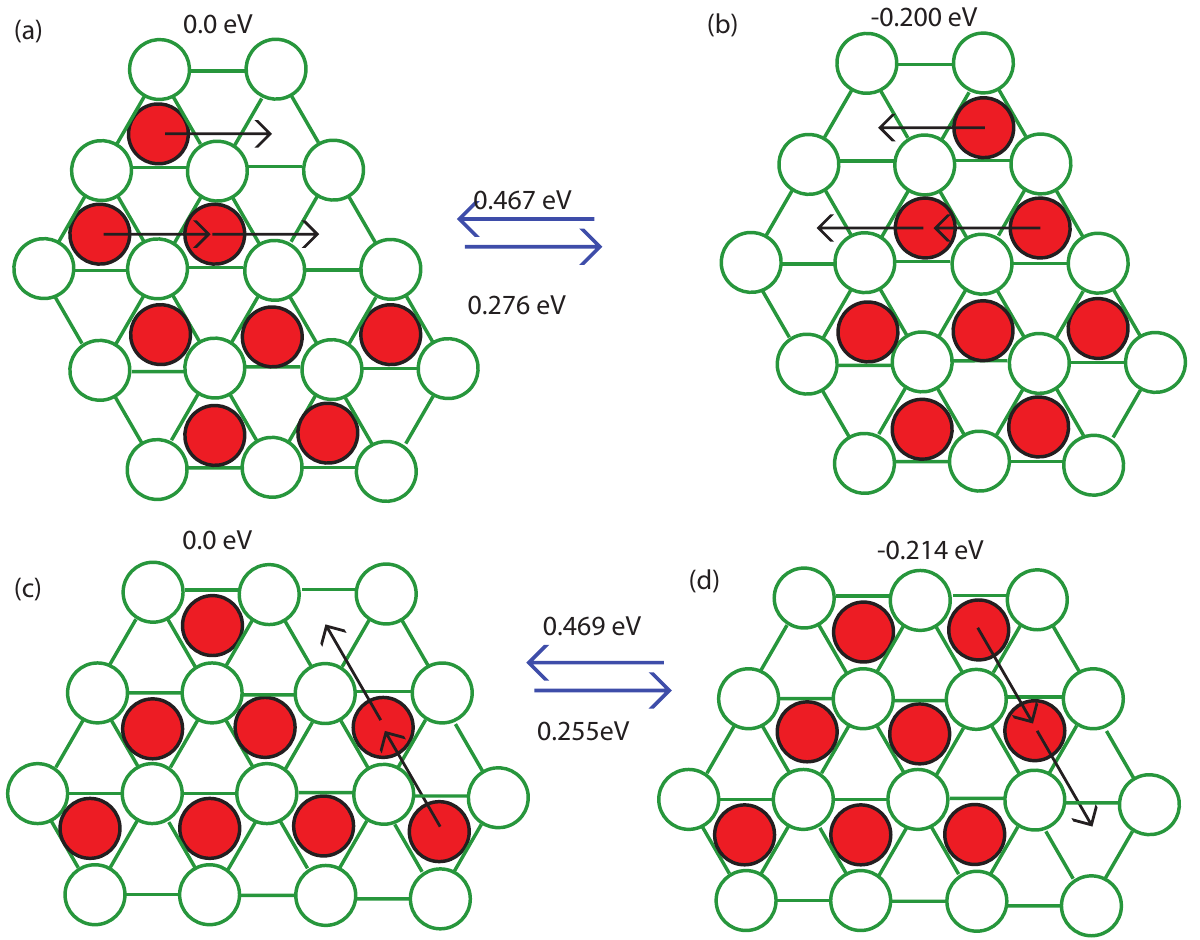} 
\caption{\label{8shearing}{Examples of shearing diffusion processes along with their activation barriers  for an 8-atom islands. (a \& b) Trimer shearing (c \& d) Dimer shearing.}}}
\end{figure}

\begin{figure}                                                                                                                                                                                                                                                                                                                                                                                                                                                                                                                                                                                                                                                                                                                                                                                                                                                                                                                                                                                                                                                                                                                                                                                                                                                                                                                                                                                                                                                                                                                                                                                                                                                                                                                                                                                                                                                                                                                                                                                                                                                                                                                                                                                                                                                                                                                                                                                                                                                                                                                                                                                                                                       
\center{\includegraphics [width=8.5cm]{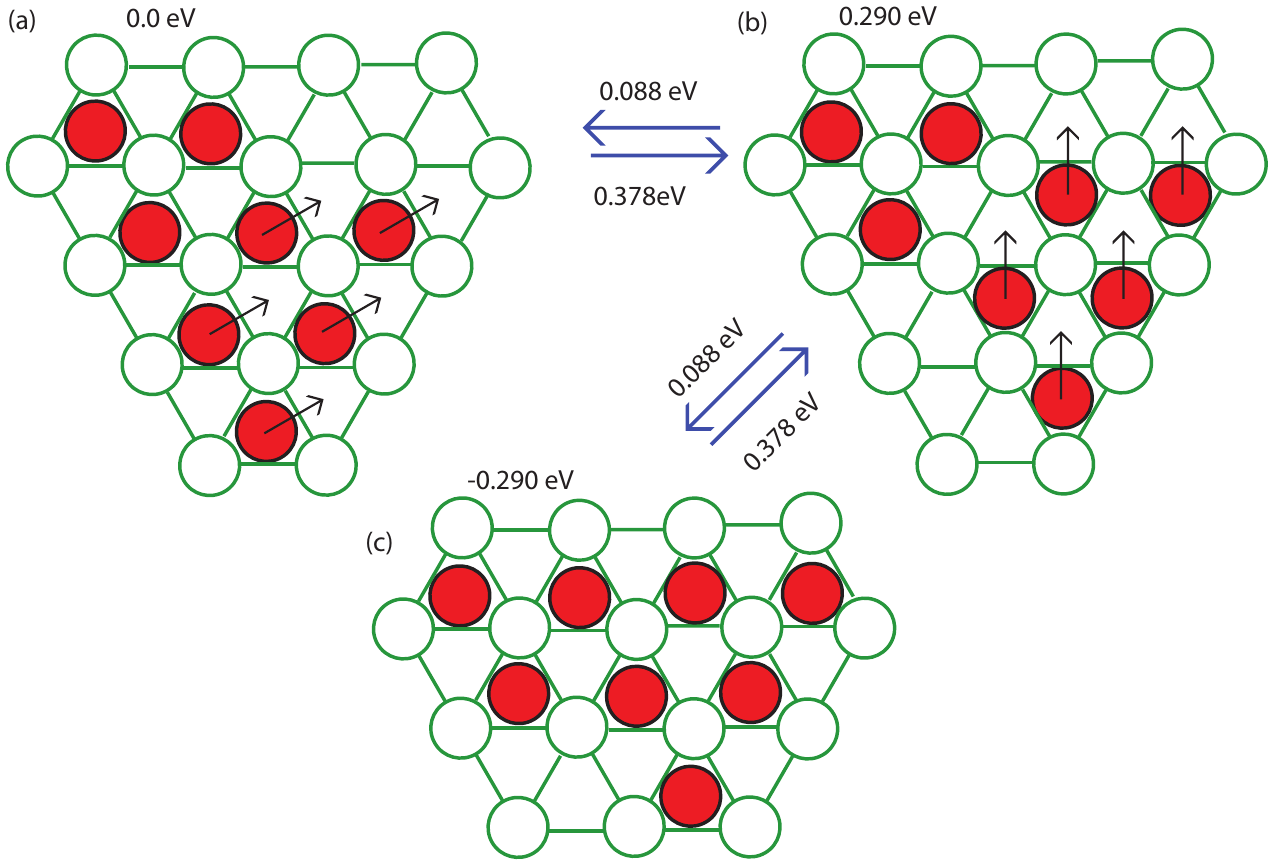} 
\caption{\label{8reptation}{Shows different processes (or step) involved in 8-atom reptation diffusion mechanism}}}
\end{figure}

Interesting multi-atom processes includes shearing and reptation as shown in Figs.~\ref{8shearing} \& \ref{8reptation} are also found in our simulations. In the case of shearing processes, part of the island (more that one atom) moves from fcc (hcp) to the nearest fcc (hcp) sites if the island is initially on fcc (hcp) sites. An example of trimer shearing process along with their activation barriers is shown in Fig.~\ref{8shearing} (a) \& (b) while Fig.~\ref{8shearing} (c) \& (d) shows a dimer shearing process and its activation barrier.

Another important multi-atom process is reptation through which an entire island diffuses from all-fcc (all-hcp) to all-hcp (all-fcc) sites in a two step shearing process. In the first step part of the island moves from fcc (hcp) to nearest hcp (fcc) sites (see Fig.\ref{8reptation}(a \& b)) and in this intermediate state, part of the island is on fcc sites and part of it is on  hcp sites. In the next step, part of the island which is on fcc (hcp) sites moves to hcp (fcc) sites (see Fig.\ref{8reptation}(a \& b)) thus completing the reptation mechanism.


\subsubsection{Nonamer}

\begin{figure}
\center{\includegraphics [width=8.5cm]{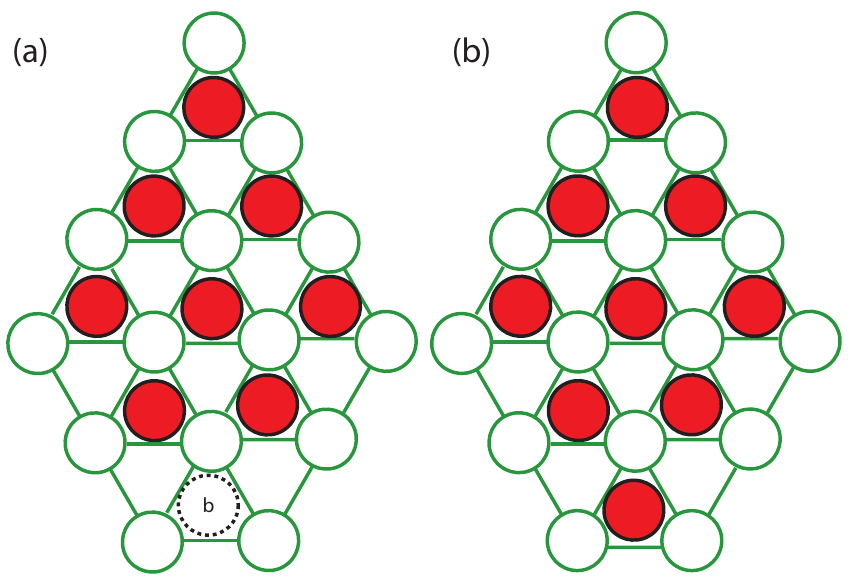} 
\caption{\label{8to9atom}{Compact diamond shaped nonamer obtained by (a) Adding an atom to the compact B-type fcc-octamer (b) Resultant compact diamond shaped fcc-nonamer.}}}
\end{figure}

\begin{figure}
\center{\includegraphics [width=9.0cm]{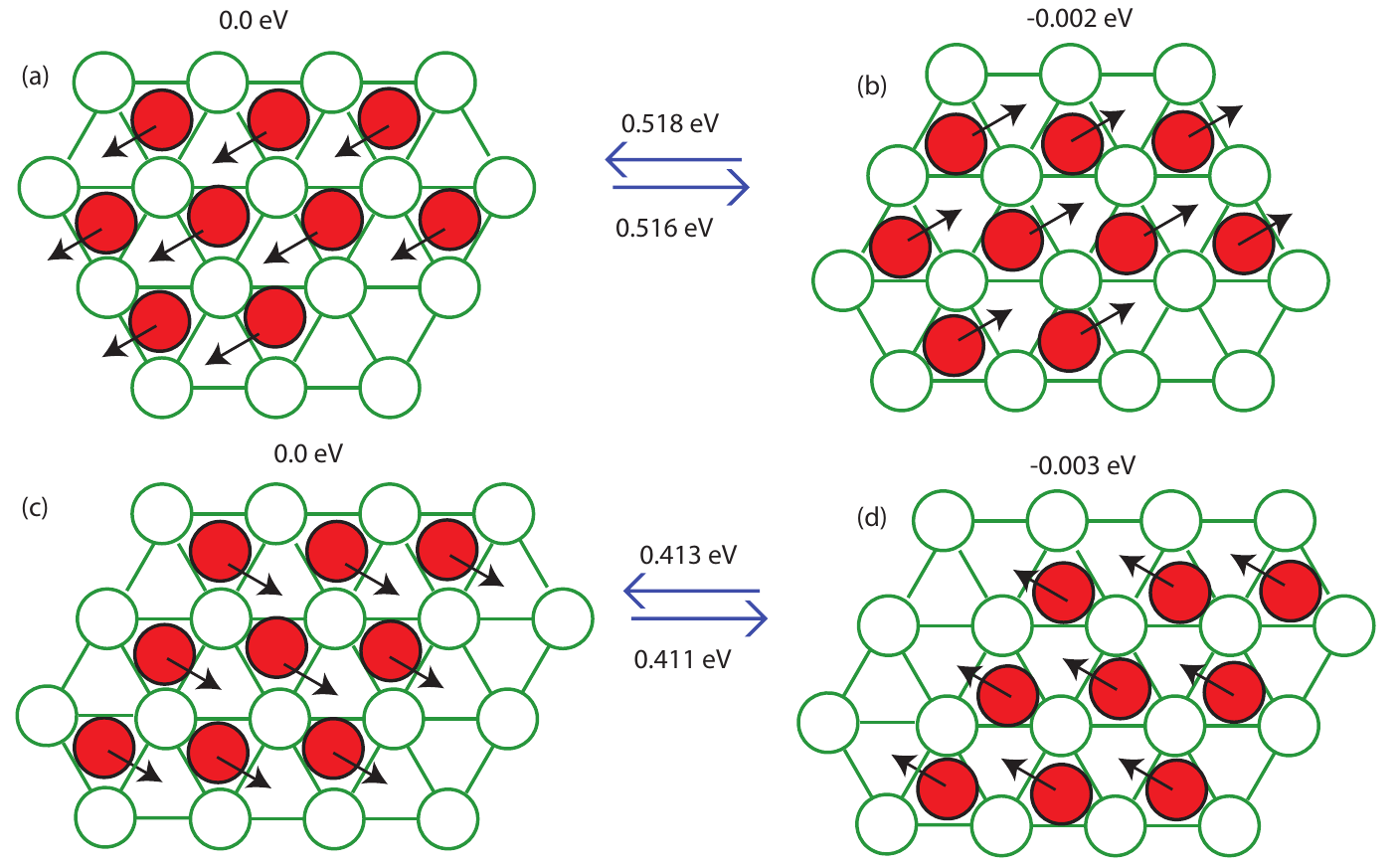} 
\caption{\label{9concerted}{Concerted diffusion processes and their activation barriers for 9-atoms compact islands.}}}
\end{figure}

\begin{table}
\caption{\label{t9concerted}Activation barriers (eV) of concerted translations processes in all $3$ directions for the 9-atom cluster shown in figs.\ref{9concerted} (a \& b, c \& d) }
\begin{tabular}{ c  c c c c c c }
\hline
\hline
Directions~~&fcc&hcp\\
\hline
\hline
1~~ & (0.518, 0.414) ~&  (0.516, 0.411) ~\\
2~~ & (0.474, 0.526) ~&  (0.472, 0.523) ~\\
3~~ &  (0.509, 0.414) ~&  (0.507, 0.411) ~\\
\hline
\hline
\end{tabular}
\end{table}

By addition of another atom to either of the compact A- or B-type fcc-octamer results in a single diamond shaped compact nonamer as shown in Fig.~\ref{8to9atom}(a - b) for the case of B-type fcc-octamer to a compact diamond shaped fcc-nonamer (having equal A- and B-step lengths). Similarly, compact A- or B-type hcp-octamer results in a single diamond shaped compact hcp-nonamer. Two other almost compact diamond shaped configurations of a nonamer for each fcc- and hcp- nonamer with a kink in A- or B-step (See Fig.~\ref{9concerted}(a - b)) are also oberved in out simulations. We find that fcc-nonamer with a kink in A-step is most stable configuration whereas fcc-nonmaer with a kink in B-step (Fig.~\ref{9concerted} (d)) is 0.002 eV higher in energy. Similarly hcp-nonamers with a kink in A-step and B-step (Fig.~\ref{9concerted} (a)) are less stable than most stable nonamer by 0.004 eV and 0.005 eV respectively. Least stable configuration are the symmetric fcc- and hcp-nonamers (Fig.~\ref{9concerted} (e \& f)) by 0.007 eV and 0.010 eV respectively.

Most frequently observed processes for a nonamer are single-atom processes like edge diffusion processes either along A- or B-type step edges or corner rounding and kink detachment processes (K3 $\rightarrow$ A2(B2), see Fig.~\ref{psingle}). In contrast, concerted diffusion processes although picked less often but does contribute most to the nonamer diffusion, since they produce large displacement in the center of mass, and as usual with these types of processes a compact island converts from an all-fcc to an all-hcp island. 
Although single-atom process individually do not produce big displacement to the center of mass of an island, in case of a nonamer they are the most frequently observed processes and do contribute to smaller extent to the island diffusion. This can be concluded by comparing  effective activation barrier for this island size in table.~\ref{teff}, and activation barriers for single atom processes in table.~\ref{tsingle} and concerted processes in table.~\ref{t9concerted}. Effective activation barrier is slightly larger than the maximum activation barrier for most frequently observed concerted processes.
Some of the most frequently observed compact nonamer shapes, corresponding concerted diffusion processes and their activation barriers are shown in Fig.~\ref{9concerted}. 
Note that energy barriers for fcc (hcp) clusters shown in figs.~\ref{9concerted}(a - d) in directions 1 and 3 have same energy barrier.
Table.~\ref{t9concerted} shows the activation barriers for concerted processes in all $3$ directions for nonamers shown in Fig.~\ref{9concerted}. Furthermore for these 9-atom islands, concerted diffusion in the direction $1$ \& $2$ are the most frequently observed concerted processes during the island diffusion.

\begin{figure}
\center{\includegraphics [width=8.0cm]{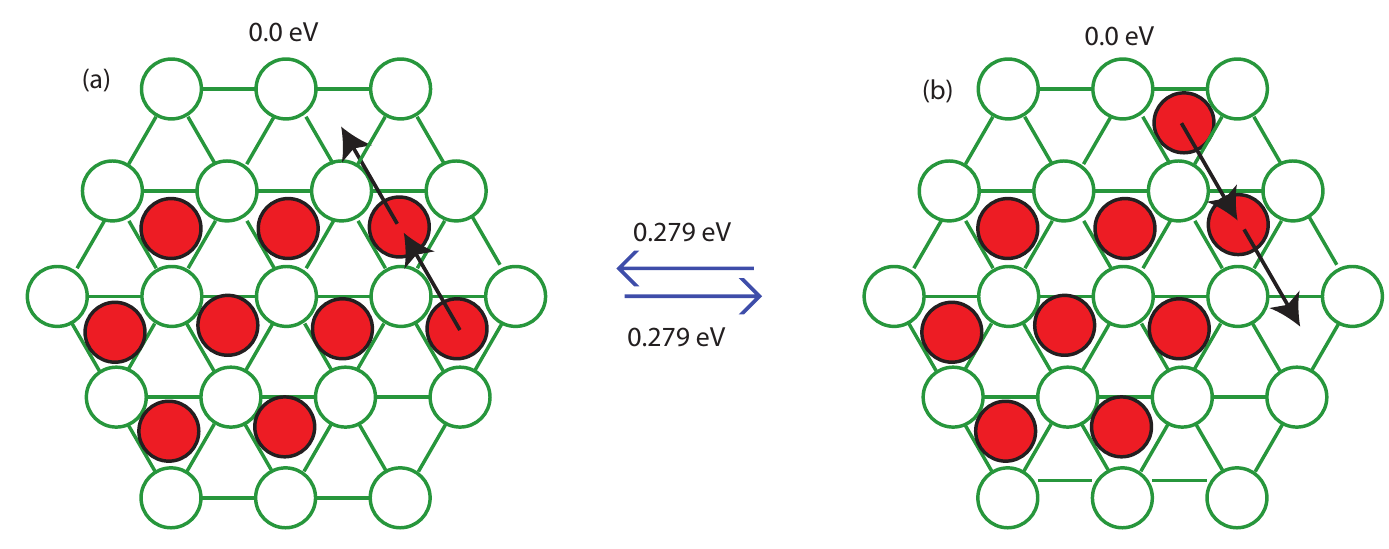} 
\caption{\label{9multi}{Dimer shearing process and their activation barriers for compact 9-atom island.}}}
\end{figure}

Fig.~\ref{9multi} shows frequently observed multi-atom process and its activation barrier during nonamer diffusion. This is again is a shearing of dimer along A-type step-edge similar to what was observed previously with other island sizes.  With this shearing process atoms diffuse from one fcc (hcp) to another fcc (hcp) sites  and appears as though a dimer is diffusion along an edge of compact 7-atom cluster. Activation barrier for this shearing process is lower than a single-atom diffusion along an edge as well as a corner rounding process. We note that this dimer forms an A-type step edge to the 9-atom island. Although not shown here, we have also observed reptation processes ~\cite{slkmcII} as well and activation barriers for the intermediate processes were ever lower than single-atom diffusion processes, but former happens only when the shape of the island is non-compact.


%

\subsubsection{Decamer}
\begin{figure}
\center{\includegraphics [width=8.5cm]{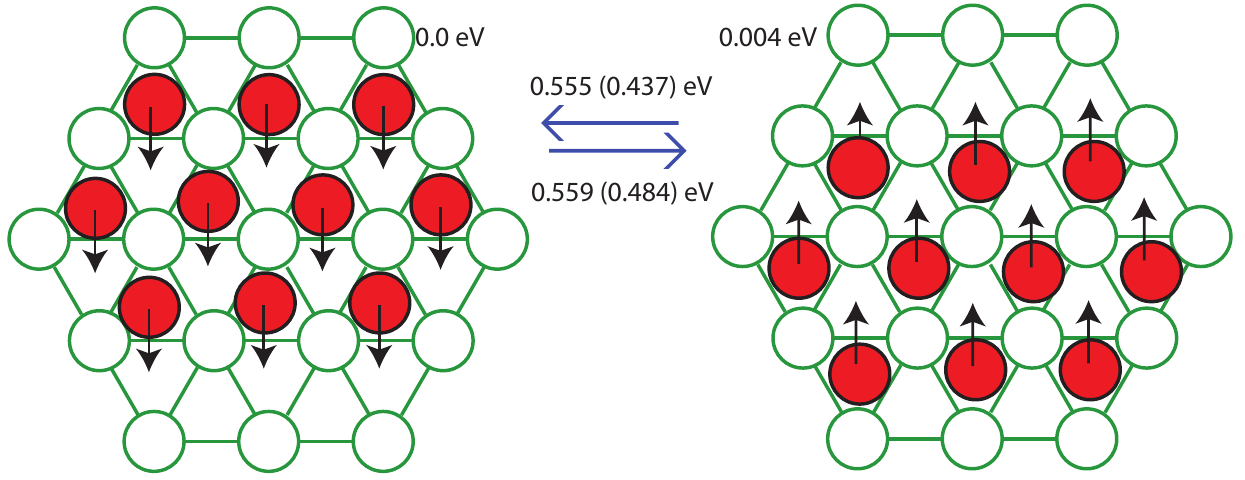} 
\caption{\label{10atoms}{Concerted diffusion processes along direction 1 for (a) fcc decamer (b) hcp decamer. Values in brackets are the energy barriers obtained from DFT calculations.}}}
\end{figure}

There are two possible compact geometries, one for each fcc- and hcp-decamer as shown in Figs.~\ref{10atoms}(a \& b) with fcc-decamer being 0.003 eV more stable than compact hcp-decamer. In the case of a decamer we have observed a variety of single-atom, multi-atom and concerted diffusion processes as well. Most frequent shaped observed in our simulations are the compact shapes as shown in Fig.~\ref{10atoms} which has same number of A- and B-type step edges of same length. For this shape most frequently observed process was concerted diffusion processes as shown in Fig.~\ref{10atoms} along with their activation barriers and as usually changes all-fcc cluster into a all-hcp cluster and vice versa. Table.~\ref{t10concerted} shows activation barriers for the concerted process shown in  Fig.~\ref{10atoms} in all $3$ directions both for fcc and hcp clusters.

\begin{table}
\caption{\label{t10concerted}Activation barriers (eV) of concerted cluster translations processes in all $3$ directions for the compact fcc and hcp decamers of Figs.~\ref{10atoms}. }
\begin{tabular}{ c  c c c c c c }
\hline
\hline
Directions~~&fcc&hcp\\
\hline
\hline
1~~ & 0.580 ~& 0.576~\\
2~~ & 0.559 ~& 0.555~\\
3~~ & 0.559 ~& 0.555~\\
\hline
\hline
\end{tabular}
\end{table}

\begin{figure}
\center{\includegraphics [width=8.5cm]{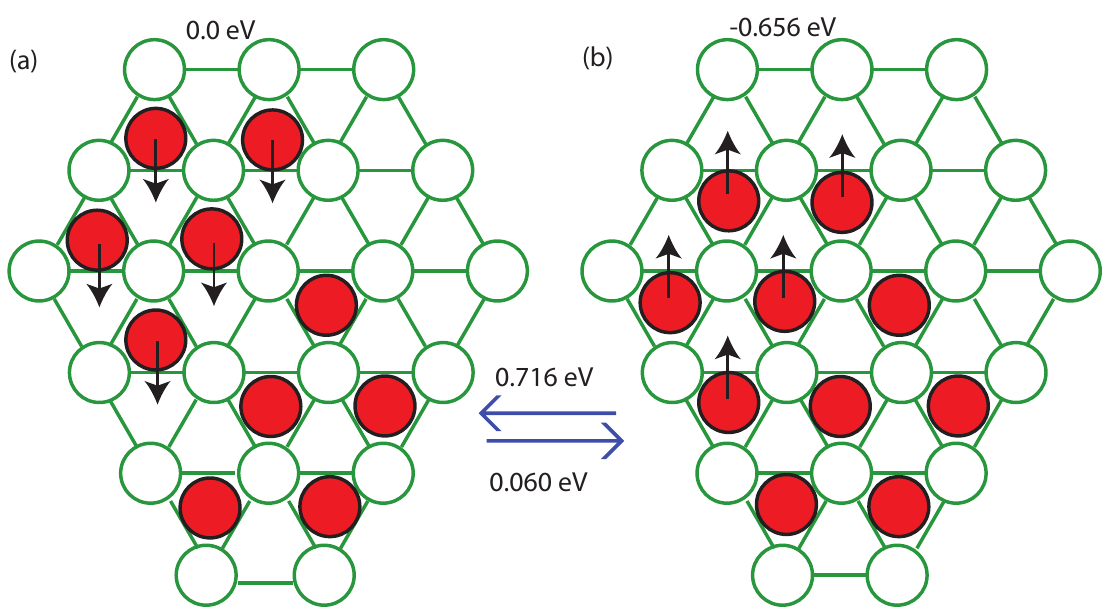} 
\caption{\label{10reptation}{A step in reptation process for a decamer along with its activation barrier. (a) five atom island moving from fcc to hcp sites. (b) its reverse.}}}
\end{figure}

For a non-compact 10-atom cluster we have also observed multi-atom processes like shearing and reptations.  Dimer shearing process along A-type step-edge similar to what has been observed for small clusters is the most frequently observed shearing process. Fig.~\ref{10reptation} shows diffusion processes involved in a reptation process along with their activation barriers

\subsubsection{Single-atom processes}\label{singleatoms}
Various types of single-atom processes like edge diffusion, corner rounding, kink attachment, kink detachment etc were automatically identified and their activation barriers were found during our simulations. Some of these single-atom processes are shown in Fig. 24(a \& b) for an hcp island with their activation barriers and those for their fcc analogs are given in Table ~\ref{tsingle}. In each single-atom process, an atom on an fcc site moves to a nearest-neighbor vacant fcc site, while an atom on hcp site moves to a nearest-neighbor vacant hcp site. These single-atom processes are also referred as long-jumps. We also found, specially for small islands of size 1-6, single-atom processes, called short-jumps, in which an atom on an fcc(hcp) site moves to nearest-neighbor hcp(fcc) site. These short-jump single-atom processes are dominant diffusion mechanism for hetero-epitaxial system with large lattice mismatch. Note that the activation barriers for single-atom processes depend not only on whether the atom is part of an fcc island or an hcp island but also on the neighborhood of the diffusing atom (like atom diffusing on an A-type or a B-type step edge, corner, kink etc).

To classify single-atom processes in Table~\ref{tsingle} we have used the notation X$_{n_{i}}$U $\rightarrow$ Y$_{n_{f}}$V, where X or Y = A (for an A-type step-edge) or B (for a B-type step-edge) or K (for kink) or C (for corner) or M (for monomer); n$_{i}$ = the number of nearest-neighbors of the diffusing atom before the process; n$_{f}$ = the number of that atom's nearest neighbors after the process. U or V = A or B (for corner or kink processes) or null (for all other other process types).

For example, process 1, B$_{2}$ $\rightarrow$ B$_{2}$, is a single-atom B-step edge process in which the diffusing atom has 2 nearest-neighbors before and after the process. Process 3, C$_{1}$B $\rightarrow$ B$_{2}$, is a corner rounding process towards a B-step, the diffusing atom starting on the corner of a B-step with one nearest-neighbor and ending up on the B-step with two nearest-neighbors. In process 10, C$_{2}$A $\rightarrow$ C$_{1}$B, the diffusing atom begins on the corner of an A-step having two nearest-neighbors and ends up on the corner of a B-step with only one nearest-neighbor.

\begin{table}
\caption{\label{tsingle}Activation barriers (eV) of single-atom processes for both fcc and hcp islands. The index numbers refer to the types of processes illustrated in Fig.~\ref{psingle}. See text for explanation of the notation used to classify the process types.}
\begin{tabular}{ c  l c c}
\hline
\hline
Index no. ~&Process type&fcc & hcp\\
\hline
\hline
1~~ &B$_{2}$ $\rightarrow$ B$_{2}$& 0.349~&0.347  ~\\
2~~ &B$_{2}$ $\rightarrow$ M&0.555 ~&0.555 ~\\
3~~ &C$_{1}$B $\rightarrow$ B$_{2}$ &0.138~&0.071  ~\\
4~~ &C$_{1}$B $\rightarrow$ M &0.319~&0.319 ~\\
5~~ &C$_{1}$B $\rightarrow$ A$_{2}$ &0.72 ~& 0.138 ~\\
6~~ &C$_{1}$B $\rightarrow$ C$_{1}$B & - ~& - ~\\
7~~ &C$_{1}$B $\rightarrow$ M & - ~& - ~\\
8~~ &A$_{2}$ $\rightarrow$ M &0.537 ~&0.537  ~\\
9~~ &A$_{2}$ $\rightarrow$ A$_{2}$ &0.266 ~&0.265  ~\\
10~~ &C$_{2}$A $\rightarrow$ C$_{1}$B &0.385 ~&0.308  ~\\
11~~ &C$_{2}$A $\rightarrow$ M &0.564 ~&0.545  ~\\
12~~ &B$_{2}$ $\rightarrow$ K$_{3}$B &0.322~& 0.322~\\
13~~ &K$_{3}$A $\rightarrow$ A$_{2}$ &0.471 ~& 0.470~\\
14~~ &A$_{2}$ $\rightarrow$ K$_{3}$A & 0.252~& 0.252 ~\\
15~~ &K$_{3}$B $\rightarrow$ B$_{2}$ &0.535  ~& 0.535  ~\\
16~~ &C$_{3}$B $\rightarrow$ C$_{1}$A &0.579 ~& - ~\\
17~~ &C$_{3}$B $\rightarrow$ M &0.753 ~&  - ~\\
18~~ &C$_{3}$B $\rightarrow$ C$_{1}$B &0.516 ~& - ~\\
\hline
\hline
\end{tabular}
\end{table}

\begin{figure}
\center{\includegraphics [width=8.5cm]{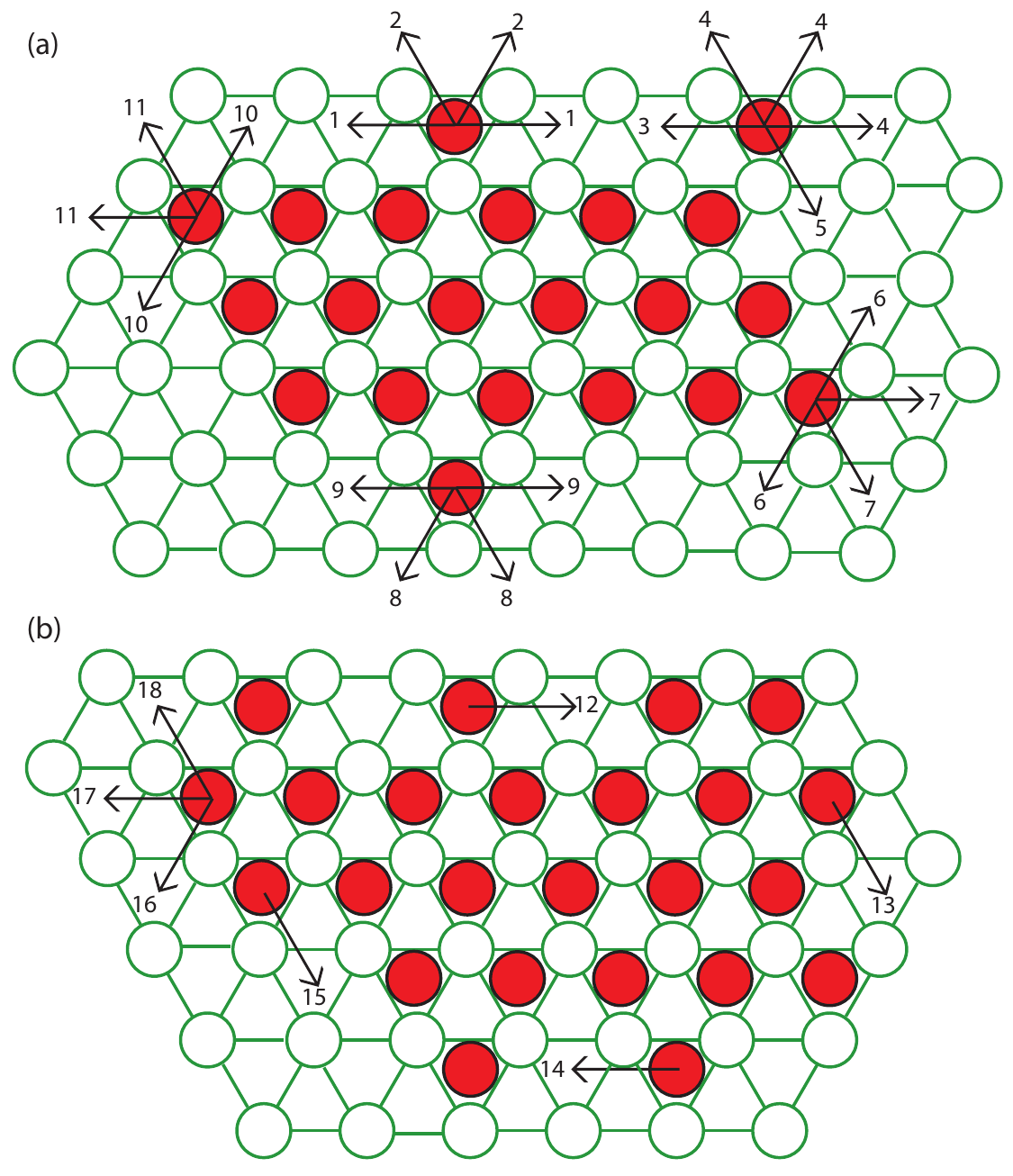} 
\caption{\label{psingle}{(Color online) Single-atom processes for an hcp island. The index numbers designate the processes described in Table.~\ref{tsingle}, which gives the activation barriers for each.}}}
\end{figure}

\section{Diffusion coefficients and effective energy barriers}\label{Diffusion coefficients and effective energy barriers}

\begin{table*}[ht]
\caption{\label{teff}Diffusion coefficients ((A$^\circ$)$^{2}$/sec) at various temperatures and effective energy barriers for Ag islands.}
\begin{tabular}{ c | c c c c c c }
\hline
\hline
Island Size &300K & 400K & 500K & 600K& 700K & $E_{eff}$ (eV)\\
\hline
\hline
1 & $2.11\times 10^{11}$~ & $3.74\times 10^{11}$ ~& $6.62\times 10^{11}$~& $5.00\times 10^{11}$~ & $7.79\times 10^{11}$ ~& 0.059 \\ 
2 & $6.33\times 10^{10}$ ~& $1.68\times 10^{11}$~& $2.76\times 10^{11}$~ & $4.13\times 10^{11}$ ~& $5.29\times 10^{11}$ ~& 0.096\\
3 & $2.94\times 10^{09}$ ~& $1.62\times 10^{10}$~& $4.72\times 10^{10}$~ & $9.51\times 10^{10}$ ~& $1.49\times 10^{11}$ ~& 0.179\\
4 & $2.20\times 10^{09}$ ~& $1.57\times 10^{10}$~& $4.48\times 10^{10}$~ & $1.14\times 10^{11}$ ~& $1.86\times 10^{11}$ ~& 0.202\\
5 & $1.08\times 10^{08}$ ~& $1.72\times 10^{09}$~& $8.73\times 10^{09}$~ & $2.83\times 10^{10}$ ~& $6.01\times 10^{10}$ ~& 0.287\\
6 & $4.50\times 10^{07}$ ~& $7.87\times 10^{08}$~& $4.36\times 10^{09}$~ & $1.45\times 10^{10}$ ~& $3.29\times 10^{10}$ ~& 0.298\\
7 & $1.14\times 10^{06}$ ~& $6.14\times 10^{07}$~& $6.90\times 10^{08}$~ & $3.07\times 10^{09}$ ~& $1.01\times 10^{10}$ ~& 0.410\\
8 & $9.55\times 10^{04}$ ~& $4.99\times 10^{06}$~& $5.64\times 10^{07}$~ & $2.81\times 10^{08}$ ~& $1.00\times 10^{09}$ ~& 0.413\\
9 & $3.48\times 10^{04}$ ~& $2.07\times 10^{06}$~& $2.51\times 10^{07}$~ & $1.90\times 10^{08}$ ~& $7.12\times 10^{08}$ ~& 0.425\\
10 & $2.15\times 10^{02}$ ~& $7.49\times 10^{04}$~& $2.51\times 10^{06}$~ & $2.57\times 10^{07}$ ~& $1.29\times 10^{08}$ ~& 0.501\\
\hline
\hline
\end{tabular}
\end{table*}

\begin{figure}[ht]
\center{\includegraphics [width=7.0cm]{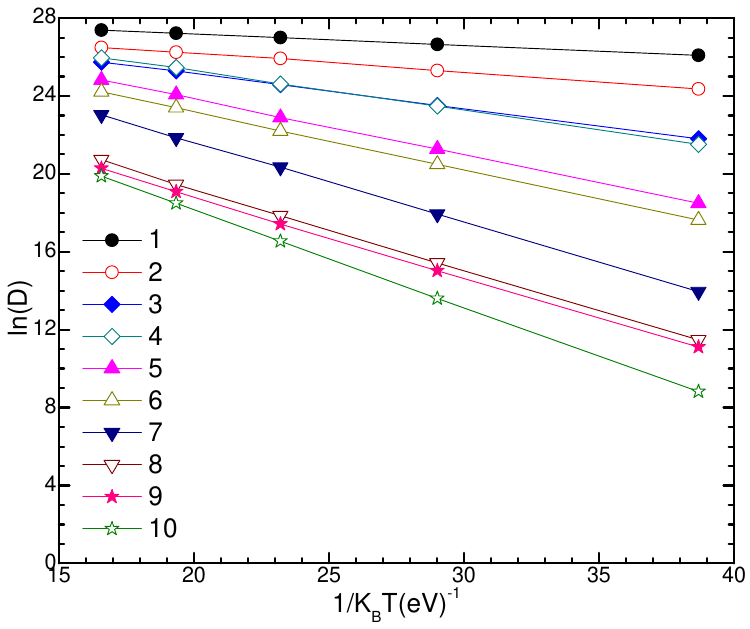} 
\caption{\label{arr}{Arrhenious plots for 1$\--$10 atom islands.}}}
\end{figure}

\begin{figure}[ht]
\center{\includegraphics [width=7.0cm]{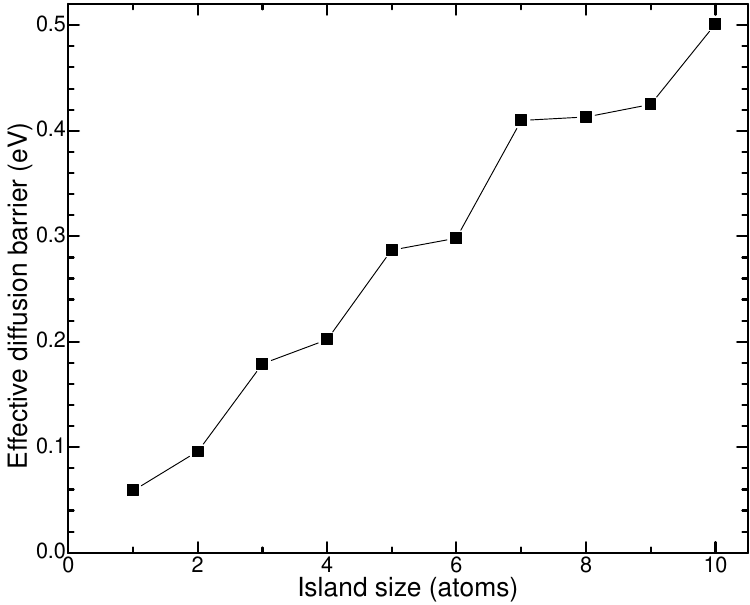} 
\caption{\label{eff}{Effective energy barriers of 1$\--$10 atom islands as a function of island size.}}}
\end{figure}

We start our simulations with an empty database. Every time a new configuration (or neighborhood) is found, SLKMC-II finds all possible processes and their activation barriers for it on-the-fly and stores them in a database. Since the processes and their activation barriers are dependent on island size, each one of them has separate unique database.
We carry out 10$^{7}$ KMC steps for each island size at temperatures 300K, 400K, 500K, 600K and 700K. The diffusion coefficient of an adatom island is calculated by D = lim$_{t \rightarrow \infty }$ ([R$_{COM}$(t) - R$_{COM}$(0)]$^{2}$)/2dt, where D is the diffusion coefficient, R$_{COM}$(t) is the position of the center of mass of the island at time t, and d is the dimensionality of the system which in our case is $2$. Diffusion coefficients thus obtained for island sizes $1-10$ are summarized in Table~\ref{teff}. 
Effective energy barriers for each island size are extracted from their respective Arrhenius plots  as shown in Fig.~\ref{arr} while Fig.~\ref{eff} shows plot of effective energy barriers as a function of island size.

\section{Discussion}\label{Discussion}
\begin{figure}
\center{\includegraphics [width=7.0cm]{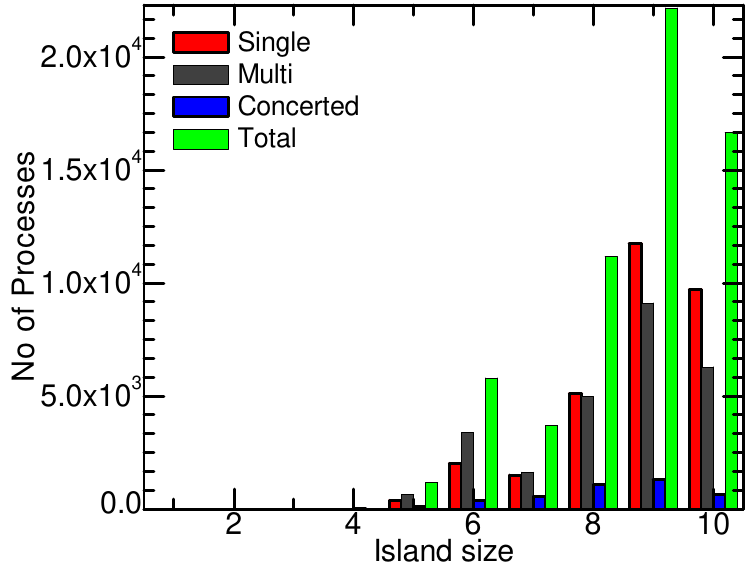} 
\caption{\label{database}{Distribution of single-atom, multi-atom, concerted and total processes for 1-10 atom islands accumulated in the database during SLKMC simulations.}}}
\end{figure}
To summarize, we have performed a systematic study of the diffusion of small Ag (1-10 atom) islands on Ag(111) surface, using newly developed self learning KMC (SLKMC-II) simulations in which the system is allowed to evolve through mechanisms of its choice with the usage of a self-generated database of single-atom, multi-atom and concerted diffusion processes. Specifically, diffusion processes like concerted processes and multi-atom processes involving fcc-fcc, fcc-hcp and hcp-hcp jumps are automatically found during the simulations. We find that these small-sized islands diffuse primarily through concerted motion with a small contribution from single atom processes, even though for certain cases the frequency of single atom processes is large because of lower activation energies. Furthermore multi-atom processes including shearing and reptation, which are activated at higher temperature, contributes rarely to the diffusion of island. Interestingly enough, energy barriers for reptation processes are small but they are only occurred when island shape is non-compact which is rare for the temperature range we study here.
We present in Fig.~\ref{database} the total number of processes found by our simulations for different island sizes. As can be seen from it, number of processes increase exponentially, specially beyond 4-atom island, suggesting the need for an automatic way of finding all the possible processes during simulations. Also by the inclusion of fcc and hcp sites in the pattern recognition scheme allows us to get all possible concerted processes which involve motion from fcc-hcp sites. Fig.~\ref{database} also shown single atom and multi atom processes.
Allowing the system the possibility of evolving in time through all types of processes of its choice, we are able to establish the relative significance of various types of atomistic processes through considerations of the kinetics and not just the energetics and/or the thermodynamics, as is often done. For small Ag islands on Ag(111), we find the effective barriers for diffusion to scale with island size.

\begin{acknowledgments}

We would like to acknowledge computational resources provided by University of Central Florida. We also thank Lyman Baker for critical reading of the manuscript.

\end{acknowledgments}

%

 \end{document}